\documentclass[fleqn,usenatbib]{mnras}

\usepackage{newtxtext,newtxmath}
\usepackage[T1]{fontenc}
\DeclareRobustCommand{\VAN}[3]{#2}
\let\VANthebibliography\thebibliography
\def\thebibliography{\DeclareRobustCommand{\VAN}[3]{##3}\VANthebibliography}
\usepackage{graphicx}	
\usepackage{amsmath}	
\usepackage[dvipsnames]{xcolor}
\usepackage{multirow}
\usepackage{booktabs}
\usepackage{calc}
\usepackage{xcolor}
\usepackage{adjustbox}
\usepackage{soul}
\newcommand{\bl}[1]{\mbox{\boldmath$ #1 $}}
\usepackage[normalem]{ulem} 

\title[Pebbles in protoplanetary discs]{Ices on pebbles in protoplanetary discs}

\author[A. Topchieva et al.]{
A. Topchieva,$^{1}$\thanks{E-mail: ATopchieva@inasan.ru}
T. Molyarova,$^{1}$
V. Akimkin,$^{1}$
L. Maksimova,$^{1}$
and E. Vorobyov$^{1,2}$
\\
$^{1}$Institute of Astronomy, Russian Academy of Sciences, 48 Pyatnitskaya St., Moscow, 119017, Russia\\
$^{2}$Research Institute of Physics, Southern Federal University, Rostov-on-Don, Russia
}

\date{Accepted XXX. Received YYY; in original form ZZZ}

\pubyear{2023}

\begin{document}
\label{firstpage}
\pagerange{\pageref{firstpage}--\pageref{lastpage}}
\maketitle

\begin{abstract}

The formation of solid macroscopic grains (pebbles) in protoplanetary discs is the first step toward planet formation. We aim to study the distribution of pebbles and the chemical composition of their ice mantles in a young protoplanetary disc. We use the two-dimensional hydrodynamical code FEOSAD in the thin-disc approximation, which is designed to model the global evolution of a self-gravitating viscous protoplanetary disc taking into account dust coagulation and fragmentation, thermal balance, and phase transitions and transport of the main volatiles (H$_2$O, CO$_{2}$, CH$_{4}$ and CO), which can reside in the gas, on small dust ($<1$\,$\mu$m), on grown dust ($>1$\,$\mu$m) and on pebbles. We model the dynamics of the protoplanetary disc from the cloud collapse to the 500\,kyr moment. We determine the spatial distribution of pebbles and  composition of their ice mantles and estimate the mass of volatiles on  pebbles, grown dust  and small dust. We show that pebbles form as early as 50\,kyr after the disc formation and exist until the end of simulation (500\,kyr), providing prerequisites for planet formation.
All pebbles formed in the model are covered by icy mantles. Using a model considering accretion and desorption of volatiles onto dust/pebbles, we find that the ice mantles on  pebbles consist mainly of H$_2$O and CO$_{2}$, and are carbon-depleted compared to gas and ices on small and grown dust, which contain more CO and CH$_4$. This suggests a possible dominance of oxygen in the composition of planets formed from pebbles under these conditions. 
\end{abstract}

\begin{keywords}
astrochemistry -- accretion, accretion discs -- protoplanetary discs -- hydrodynamics --  solid state: volatile -- stars: pre-main-sequence
\end{keywords}

\section{Introduction}

There is a number of unsolved problems in the theory of planet formation, e.g., the formation of kilometre-scale planetesimals from micron grains. In the context of giant planet formation, the subsequent stages may possibly involve the core accretion mechanism \citep{1980PThPh..64..544M,1986Icar...67..391B,1996Icar..124...62P,2009ApJ...695L..53B,2014prpl.conf..643H}. Recently,  more attention is drawn to the mechanism of pebble accretion \citep{Ormel2010, Lambrechts2012, 2015Icar..258..418M, Ida2016A&A, Lambrechts2017, 2017AREPS..45..359J}, 
which may solve many problems related to the timescale of the gas and ice giants formation \citep{2015A&A...582A.112B}. Given the relatively short lifetimes of protoplanetary discs \citep[on the order of $5-10$ million years;][]{2001ApJ...553L.153H,2009AIPC.1158....3M,2014ApJ...793L..34P}, pebbles may contribute significantly to planet formation.

Pebbles are macroscopic solid dust grains, typically larger than a few millimetres, that  are large enough to drift relative to gas (and to small dust grains dynamically coupled to gas), but small enough for gas to have a noticeable effect on their dynamics in the conditions of protoplanetary discs \citep{Lambrechts2012}. Because of friction with the gas, pebbles accrete onto protoplanetary cores much more efficiently than planetesimals, thus accelerating the process of planet formation \citep{Ormel2010}. The chondrules present in primitive meteorites have characteristic sizes on the order of millimetres, indicating that their parent bodies may have formed as a result of pebble accretion~\citep{2015SciA....1E0109J}. The fragmentation barrier and dust drift barrier also give a similar value of maximum dust size in protoplanetary discs~\citep[from millimetres to decimetres,][]{2014prpl.conf..339T}.

The collisional evolution of dust and its interaction with gas allow dust grains to grow to macroscopic sizes ($\sim$1\,cm), but this process does not lead directly to the formation of planetesimals~\citep{2014prpl.conf..339T}. Besides coagulation, which leads to grains sticking and growing in size, dust is also subject to fragmentation, if the grains collide with impact velocities exceeding some threshold value $v_{\rm frag}\sim1-10$\,m~s$^{-1}$. The dust size distribution function is the result of the balance of these two processes, and the fragmentation velocity $v_{\rm frag}$ plays an important role in establishing this balance.

The studies by \citet{2020A&A...637A...5E} and \citet{Vorobyov2023A&A} show that the amount and spatial distribution of pebbles in the protoplanetary disc depend significantly on the fragmentation velocity.

The fragmentation velocity, in turn, depends on the properties of dust grains, in particular, on the presence of ice mantles that weakens the fragmentation~\citep{2009ApJ...702.1490W,2015ApJ...798...34G,2016ApJ...821...82O}. Therefore, ice mantles on the dust grain surface should favour the formation of pebbles. Destruction of ice mantles by luminosity outbursts characteristic of young stars leads to potentially observable changes in the dust properties in protoplanetary discs~\citep{Vorobyov2022,2023MNRAS.521.5826H}. The distribution of volatiles in the disc determines the chemical composition of the ice mantles and changes dramatically in the vicinity of the snowlines. In particular, various ices accumulate at their respective snowlines~\citep{1988Icar...75..146S,2017A&A...600A.140S,2018ApJ...864...78K,Molyarova2021ApJ}. Snowlines are also associated with the formation of planetesimals~\citep{2017A&A...608A..92D,2012ApJ...752..106O}.

The chemical composition of the ice covering the dust grains, and particularly pebbles, is noteworthy because it may be related to the future composition of the proto-atmospheres of the forming planets~\citep{2011ApJ...743L..16O, 2018A&A...613A..14E, 2019AJ....158..194O, 2021ExA....51..661L, 2022A&A...667A.160E}. Throughout its evolution, the protoplanetary disc is affected by many physical and chemical factors and undergoes significant changes, such as episodic accretion outbursts~\citep[see][]{1998ApJ...495..385H,2014prpl.conf..387A}. However, to date, there are not many studies on the evolution of the chemical composition of pebbles subject to collisional evolution in protoplanetary discs. Few mechanisms have been found to explain the possible change in the chemical composition of pebbles with time. For example, there is an effect of diffusion of vaporised ice mantles back behind the snowline~\citep[so-called cold finger effect][]{2009ApJ...704.1471M}, as well as a similar effect related to azimuthal variations of gas and dust radial velocities in a self-gravitating protoplanetary disc~\citep{Molyarova2021ApJ}.

In the work by \citet{Vorobyov2023A&A} it was shown that pebbles are present in the disc from the early stages of its evolution, but the ice mantles on dust grains were not taken into account to determine the fragmentation efficiency. Modelling of the dynamics of multicomponent dust taking into account the chemical composition (
although without changes in dust composition due to sublimation or freeze-out) in the three-dimensional SPH model was carried out in \citet{2017MNRAS.469..237P}. It was shown that the drift of dust grains is affected by their material density, leading to the segregation of dust grains of different chemical composition. Using a 1D viscous disc model with dust evolution and chemical kinetics, \citet{2019MNRAS.487.3998B} showed that dust radial drift leads to the enhancement of molecular abundances inside their snowlines. Their model also accounted for the difference in $v_{\rm frag}$ between ice-covered and silicate grains.

In this paper, we study the distribution of pebbles in the protoplanetary disc and the composition of their ice mantles using the 2D thin-disc FEOSAD code~\citep{2018A&A...614A..98V}. Modelling of advection and adsorption/desorption of volatiles is based on the model from \citet{Molyarova2021ApJ}, and the definition of pebbles is adopted from \citet{Vorobyov2023A&A} with some modifications. The paper is organised as follows. In Section~\ref{sec:results1} we describe the numerical model and its parameters, emphasising new features compared to the \cite{Vorobyov2018A&A...614A..98V}: the definition of pebbles (Section~\ref{sec:2B}) and the volatiles evolution model (Section~\ref{sec:chem_model}). In Section~\ref{sec:res}, we present the evolution of the composition of the solid component, including ice mantles on small and grown dust and on pebbles. We also study an impact of the initial cloud mass in Section~\ref{sec:mass}. In Section~\ref{sec:obs}, we discuss individual model parameters and their influence on the properties of pebbles in the disc the variation of the initial cloud mass. In Section~\ref{sec:vivod}, we make conclusions about the dynamics of pebbles in the protoplanetary disc and the composition of their ice mantles.

\section{Model description}
\label{sec:results1}

In this paper, we use numerical simulations of the formation and evolution of the protostellar/protoplanetary disc using the FEOSAD (Formation and Evolution Of Stars And Discs) model presented in \citet{Vorobyov2018A&A...614A..98V}, with further modifications in \citet{Molyarova2021ApJ}. FEOSAD is a two-dimensional hydrodynamic code in a thin-disc approximation. The simulation starts from the gravitational collapse of a prestellar cloud and ends at the T~Tauri stage. The simulation captures spatial scales from a few to thousands of au with a spatial resolution on the log-spaced numerical grid that allows us to study the evolution of various structures in the disc, such as spiral arms, rings, and self-gravitating clumps. 

One of the advantages of such 2D modelling is that it uses the full set of hydrodynamic equations and at the same time does not require significant computational resources compared to the full 3D approach. 
It also allows to more accurately calculate thermal balance as it employs time-dependent energy equation instead of polytropic approximation frequently used in 3D models. A natural shortcoming of the 2D thin disc approximation is that it considers vertically integrated surface densities and lacks the detailed description of the disc vertical structure. For example, the snowlines of the volatiles in the upper layers can be situated at different distances than those in the midplane. It also omits some dynamical processes for which vertical direction is essential, such as meridional circulation.
This kind of a thin-disc approach is a useful tool to study the long-term evolution of protoplanetary discs.

The FEOSAD code includes the following physical processes: self-gravity of the disc, turbulent viscosity parameterised using the $\alpha$ approach of \cite{Shakura1973A&A}, radiative cooling and heating by the stellar and background irradiation, viscous heating, dust evolution, dust drift relative to gas, including back reaction of dust on gas, and 
phase transitions and transport of the main volatiles. Dust evolution is determined by collisional growth and fragmentation. Friction between dust and gas includes back reaction of dust grains on gas (see Section~\ref{sec:dust_model} for more details).
Central star properties evolve with time according to stellar evolution code STELLAR~\citep{2008ASPC..387..189Y,2017A&A...605A..77V} depending on gas accretion to the central sink cell.  

\subsection{Dynamics of gas and dust}
\label{sec:dust_model}

Hydrodynamic equations describing the conservation laws for mass, momentum and energy for the gas component have the form:

\begin{equation}
\frac{{\partial \Sigma_{\rm g} }}{{\partial t}} + \nabla_p \cdot 
\left( \Sigma_{\rm g} \bl{v}_p \right) =0, 
\label{eq:1}
\end{equation}

\begin{eqnarray}
\frac{\partial}{\partial t} \left( \Sigma_{\rm g} \bl{v}_p \right) + [\nabla \cdot \left( \Sigma_{\rm
g} \bl{v}_p \otimes \bl{v}_p \right)]_p & =& - \nabla_p {\cal P} + \Sigma_{\rm g} \, \bl{g}_p + \nonumber
\\ 
+ (\nabla \cdot \mathbf{\Pi})_p - \Sigma_{\rm d,gr} \bl{f}_p,
\label{eq:2}
\end{eqnarray}

\begin{equation}
\frac{\partial e}{\partial t} +\nabla_p \cdot \left( e \bl{v}_p \right) = -{\cal P} 
(\nabla_p \cdot \bl{v}_{p}) -\Lambda +\Gamma + 
\left(\nabla \bl{v}\right)_{pp^\prime}:\Pi_{pp^\prime}, 
\label{energ}
\end{equation}
where the lower indices $p$ and $p^\prime$ denote the planar components ($r$, $\phi$) in polar coordinates, $\Sigma_{\rm g}$ is the gas surface density, $e$ is the internal energy per unit surface area, $\cal P$ is the vertically integrated gas pressure determined from the ideal equation of state ${\cal P}=(\gamma-1) e$ with $\gamma= 7/5$, $\bl{f}_{p}$ is the friction force between gas and dust \citep[see][for more details]{Vorobyov2023A&A}, $\bl{v}_{ p} = v_{r}\bl{\hat{r}}+v_{\phi}\bl{\hat{\phi}}$ is the gas velocity in the disc plane, $\nabla_{p} = \hat{\bl r} \partial / \partial r + \hat{\bl \phi}r^{-1} \partial / \partial \phi$ is the gradient along the planar coordinates of the disc. The gravitational acceleration in the disc plane is defined as $\bl g_{p} = g_{r}\hat{\bl r} + g_{\phi}\hat{\bl \phi}$ and includes the gravity of the central protostar and the self-gravity of gas and dust in the disc, which is calculated by solving the Poisson integral and $\mathbf{\Pi}$ is the viscous stress tensor~\citep{Vorobyov2010ApJ...719.1896V}. Expressions for radiative cooling $\Lambda$ and heating $\Gamma$ (the latter includes stellar and background irradiation) can be found in \cite{Vorobyov2018A&A}. The background radiation temperature $T_{\rm bg}$ is assumed to be 15\,K.

Gas and dust evolve together and interact with each other through gravity and friction force, which accounts for the back reaction of dust on gas according to the analytical method described in \citet{Stoyanovskaya2018ARep,2014MNRAS.443..927L}. Initially, all dust is in the form of small submicron grains, but it grows as the protoplanetary disc forms and evolves. The small submicron dust is assumed to be coupled to the gas, while the grown dust can be dynamically separated from the gas. In this paper, we adopt the same dust growth model as described in \cite{Molyarova2021ApJ}. In the FEOSAD code, the small and grown dust populations are represented through their surface densities $\Sigma_{\rm d,sm}$ and $\Sigma_{\rm d,gr}$, respectively. Each dust population has a size distribution according to the power law $N(a) = Ca^{-p}$ with the normalisation constant $C$ and fixed exponent $p = 3.5$. For small dust population, the minimum size is $a_{\rm min}= 5 \times 10^{-7}$\,cm, and the maximum size is $a_{\rm *} = 10^{-4}$\,cm. For grown dust, $a_{\rm *}$ is the minimum size, and $a_{\rm max}$ is the maximum size, which varies in space and time due to dust coagulation and fragmentation. These distributions are similar to the classic interstellar dust size distribution of \citet{1977ApJ...217..425M}, albeit with different size limits.

The dynamics of small and grown dust grains is described by the following continuity and momentum equations (note that small dust is assumed to be strictly dynamically coupled to gas):

\begin{equation}
\label{contDsmall}
\frac{{\partial \Sigma_{\rm d,sm} }}{{\partial t}}  + \nabla_p  \cdot 
\left( \Sigma_{\rm d,sm} \bl{v}_{ p} \right) = - S(a_{\rm max}), 
\end{equation}

\begin{equation}
\label{contDlarge}
\frac{{\partial \Sigma_{\rm d,gr} }}{{\partial t}}  + \nabla_p  \cdot 
\left( \Sigma_{\rm d,gr} \bl{u}_{ p} \right) = S(a_{\rm max}) + \nabla  \cdot \left( D \Sigma_{\rm g} \nabla\left(\frac{\Sigma_{\rm d,gr}}{\Sigma_{\rm g}} \right)\right),  
\end{equation}

\begin{eqnarray}
\label{momDlarge}
\frac{\partial}{\partial t} \left( \Sigma_{\rm d,gr} \bl{u}_{ p} \right) +  \left[\nabla \cdot \left( \Sigma_{\rm
d,gr} \bl{u}_{ p} \otimes \bl{u}_{p} \right)\right]_{ p}  &=&   \Sigma_{\rm d,gr} \, \bl{g}_{p} + \nonumber \\
+ \Sigma_{\rm d,gr} \bl{f}_{ p} + S(a_{\rm max}) \bl{v}_{p},
\end{eqnarray}
where $\bl u_{\rm p}$ is the grown dust velocity. The term $S(a_{\rm max})$ is the conversion rate of small dust into the grown dust per unit surface area (in~g~s$^{-1}$~cm$^{-2}$). The last term in the equation~\eqref{contDlarge} is responsible for the turbulent diffusion of dust, similar to \citet{2020A&A...643A..13V}. The turbulent diffusivity $D$ is related to the kinematic viscosity $\nu$ as $D=\nu/\mathrm{Sc}$, with the Schmidt number $\mathrm{Sc}=1$.

In the dust growth model, the maximum grown dust size $a_{\rm max}$ is calculated at each time step for each cell as a solution to the following equation:
\begin{equation}
{\frac{\partial a_{\rm max}}{\partial t}} + (u_{\rm p} \cdot \nabla_p ) a_{\rm max} = \cal{D},
\label{dustA}
\end{equation}
where the dust growth rate $\cal{D}$ describes the change in $a_{\rm max}$ due to coagulation and fragmentation, and the second term on the left-hand side describes the change in $a_{\rm max}$ due to dust advection. We calculate the dust growth rate $\cal{D}$ as
\begin{equation}
\cal{D}=\frac{\rho_{\rm d} {\it v}_{\rm rel}}{\rho_{\rm s}},
\label{GrowthRateD}
\end{equation}
where $\rho_{\rm d}$ is the volume density of dust in the midplane, $\rho_{\rm s}=3$\,g~cm$^{-3}$ is the density of the refractory core of dust grains, and $v_{\rm rel}$ is the collision velocity between dust grains. The source term $\cal{D}$ describes dust evolution due to coagulation and fragmentation in the collision between dust grains of both populations. Eq.~\eqref{GrowthRateD} is based on the monodisperse model of \citet{1997A&A...319.1007S}, adapted for our two-population case by using total dust volume density instead of only that of grown dust.  We assume that Brownian motion and turbulence are the main sources of collisions. Thus relative velocity of the colliding dust grains can be expressed from the velocity of Brownian particles $v_{\rm th}$ and turbulence-induced velocity $v_{\rm turb}$ \citep{2007A&A...466..413O,2012A&A...539A.148B}
\begin{equation}
v_{\rm rel} = \left(v_{\rm th}^2 + v_{\rm turb}^2\right)^{1/2}
= \left(\frac{16 k_{\rm B} T_{\rm mp}}{\pi m_{\rm a}} + 3 \alpha_{\rm eff} \mathrm{St} c_{\rm s}^2\right)^{1/2}.
\label{eq:vels}
\end{equation}
Here $k_{\rm B}$ is the Boltzmann constant, $T_{\rm mp}$ is the midplane temperature, $m_{\rm a}$ is the mass of particles of the maximum size $a_{\rm max}$, $\alpha_{\rm eff}$ is the viscosity parameter (see Section~\ref{sec:visc_model}), $\mathrm{St}$ is the Stokes number, and $c_{s}$ is the sound speed. Growth rate $\cal{D}$ depends on $a_{\rm max}$ through $m_{\rm a}=4/3\pi\rho_{\rm s} a_{\rm max}^3$ and $\mathrm{St}\sim a_{\rm max}$ (see Section~\ref{sec:2B}).
This approach is described in more detail in \citet{Vorobyov2018A&A...614A..98V} with a modification of the power-law distribution treatment presented in \citet{Molyarova2021ApJ, Vorobyov2022}. 

The value of $a_{\rm max}$ is limited by the fragmentation barrier~\citep{2012A&A...539A.148B}, the maximum dust size cannot exceed the value $a_{\rm frag}$ determined by the properties of dust grains. If at any point $a_{\rm max}$ is higher than $a_{\rm frag}$, $\cal{D}$ is set to zero. The maximum size of grown dust allowed by fragmentation is defined as
\begin{equation}
 a_{\rm frag}=\frac{2\Sigma_{\rm g}v_{\rm frag}^2}{3\pi\rho_{\rm s}\alpha c_{\rm s}^2},
 \label{eq:afrag}
\end{equation}
where $v_{\rm frag}$ is the fragmentation velocity, which depends on the presence or absence of ice mantles (see Section~\ref{sec:chem_model}).

\subsection{Viscosity parameterisation}
\label{sec:visc_model}

To calculate the turbulent viscosity the model uses a variable $\alpha$-parameter in the parameterisation by~\citet{Shakura1973A&A}. The viscosity required for mass and angular momentum transport in protoplanetary discs is considered to be primarily due to the turbulence generated by the magnetorotational instability \citep{Balbus1991ApJ, Turner2014prpl}. In FEOSAD, turbulent viscosity is accounted for through the viscous stress tensor $\Pi$ (see Eqs.~\eqref{eq:2},~\eqref{energ}), and its magnitude is defined as $\nu = \alpha c_{s} H$, where $H$ is the vertical scale height of the gas disc calculated assuming local hydrostatic equilibrium. To simulate accretion in the layered-disc model of~\cite{2010apf..book.....A} we consider an effective and adaptive parameter $\alpha_{\rm eff}$ \citep[eq. (12) in][]{Kadam2022MNRAS}. The parameter $\alpha_{\rm eff}$ is calculated as a weighted average value for the combination of the layers with active and suppressed magneto-rotational instability:
\begin{equation}
    \alpha_{\rm eff}=\frac{\Sigma_{\rm MRI}\alpha_{\rm MRI}+\Sigma_{\rm dz}\alpha_{\rm dz}}{\Sigma_{\rm MRI}+\Sigma_{\rm dz}}.
\end{equation}

The active layer is assumed to have a surface density of $\Sigma_{\rm MRI}=100$\,g~cm$^{-2}$, and the dead zone is assumed to have a surface density of $\Sigma_{\rm dz}=\Sigma_{\rm g}-\Sigma_{\rm MRI}$. We assume that in the active layer $\alpha_{\rm MRI}=10^{-3}$ \citep{2017ApJ...843..150F,2018ApJ...856..117F}.
In the dead zone, $\alpha_{\rm dz}=10^{-5}$ if the gas temperature is below $1300$\,K. Otherwise, MRI burst is developed and $\alpha_{\rm dz}$ is set to 0.1, as suggested by the results of 3D simulations of the MRI bursts~\citep{2020MNRAS.495.3494Z}. This parameterisation of $\alpha$ allows to account for the suppression of turbulence and the corresponding decrease of mass transport through the dead zone. A more detailed description is presented in \citet{Bae2014ApJ, Kadam2019ApJ, Kadam2022MNRAS}. 

\subsection{The definition of pebbles}
\label{sec:2B}

The Stokes number in protoplanetary discs is commonly used to describe dust dynamics. It is defined as
\begin{equation}
\mathrm{St} = \frac{{ \Omega_{\rm k}  \rho_{\rm s} a_{\rm max}}}{{ \rho_{\rm g} c_{\rm s}}}
\end{equation}
where $\Omega_{\rm k}$ is the Keplerian angular velocity and $\rho_{\rm g} = \Sigma_{\rm g} / \sqrt{\rm 2 \pi} H$ is the gas volume density in the midplane. Grains with $\mathrm{St}\sim1$ are most strongly affected by the gas and they drift rapidly in the radial direction toward pressure maxima. 
Millimetre- to centimetre-sized dust grains known as pebbles are characterised by a relatively high value of $\mathrm{St}$ and drift relative to the gas, in contrast to smaller dust grains coupled to gas and moving with it.

In this paper we adopt the definition of pebbles from \citet{Vorobyov2023A&A}, with some modifications. In FEOSAD, two dust populations are considered, small ($a<1$\,µm) and grown ($a>1$\,µm) dust grains. We define pebbles as the grown dust grains that have Stokes number and the dust radius above some threshold values $\mathrm{St}>\mathrm{St}_{\rm 0}$ and $a >a_{\rm peb,0}$. In regions of the disc where both of these conditions are met, the pebbles have a size distribution that is part of the size distribution of grown dust, with sizes ranging from $a_{\rm peb,min}$ to $a_{\rm max}$. The minimum pebble size $a_{\rm peb,min}$ is
\begin{equation}
    a_{\rm peb,min}=
\begin{cases}
a_{\rm St_{\rm 0}}, & \text{ if } a_{\rm St_{\rm 0}} > a_{\rm peb,0}; \\
a_{\rm peb,0}, & \text{ if } a_{\rm St_{\rm 0}} \leq a_{\rm peb,0}.
\end{cases}
\label{eq:apebmin}
\end{equation}
Here $a_{\rm St_{\rm 0}}$ is the size of dust grains that have Stokes number equal to $\mathrm{St}_{\rm 0}$ for the local disc conditions. In the disc regions where pebbles exist, this value is defined as
\begin{equation}
    a_{\rm St_{\rm 0}}=a_{\rm max}\frac{\mathrm{St}_{\rm 0}}{\mathrm{St}}.
\end{equation}

In the original definition by~\citet{Vorobyov2023A&A}, some regions of the disc with $\mathrm{St}>\mathrm{St}_{\rm 0}$ were discarded due to the restriction on the minimum pebble size. Some grown dust grains present in such regions were not considered as pebbles because the definition discarded all grown grains at the location instead of only those
with radius from $a_{\rm St_{\rm 0}}$ to $a_{\rm peb,min}$. Our modified definition includes all grown grains with size and Stokes number above the specified boundary values $\mathrm{St}_{\rm 0}$ and $a_{\rm peb,0}$ into definition of pebbles. More detailed comparison of the definition of pebbles in~\citet{Vorobyov2023A&A} and in the present work can be found in Appendix~\ref{sec:pebble_appendix}.

For the main model, we adopt the boundary values $\mathrm{St}_{\rm 0}=0.01$ and $a_{\rm peb,0}=0.05$\,cm. To test the sensitivity of the results to the definition of pebbles, we also consider the boundary values $\mathrm{St}_{\rm 0}=0.05$ and $a_{\rm peb,0}=0.25$\,cm in various combinations (see Section~\ref{sec:criteria}). The criteria for pebble definition are based on studies of \cite{Lambrechts2012, Lenz2019ApJ} and references therein. The Stokes number allows us to identify large grains moving independently of the gas, but it also depends on the local gas density. In our simulation there is an infalling envelope surrounding the disc, where $\mathrm{St}$ reaches large values even for micron-sized dust grains. The condition $a_{\rm max}>a_{\rm peb,0}$ allows us to leave them out of consideration.

The surface density of pebbles $\Sigma_{\rm peb}$ inside each cell is calculated under the assumption that its size distribution function continues with the same slope as the size distribution function for grown dust ($p = 3.5$), which leads to the following expression:
\begin{equation}
\Sigma_{\rm peb} = \frac{{ \Sigma_{\rm d, gr} \left({\sqrt{a_{\rm max}} - {\sqrt{a_{\rm peb, min}}}} \right) }}{{\sqrt{a_{\rm max}} - {\sqrt{a_{\rm *}}}}},
\label{eq:sigmapeb}
\end{equation}
where $a_{\rm *} = 1$\,µm. Pebbles defined in this way are part of the grown dust population. The surface densities of ices on pebbles are denoted as $\Sigma^{\rm peb}_i$, where $i=$~H$_2$O, CO$_{2}$, CH$_4$ or CO. They are determined from the surface densities of ices on grown dust $\Sigma^{\rm gr}_i$ assuming that mass fraction of ice on pebbles is equal to mass fraction of ice on grown dust.

\begin{equation}
\Sigma^{\rm peb}_i = \Sigma^{\rm gr}_i
\frac{\Sigma_{\rm peb}}{\Sigma_{\rm d,gr}}.
\label{eq:sigmaicepeb}
\end{equation}

\subsection{The chemical model}
\label{sec:chem_model}

We consider four volatile species, H$_2$O, CO$_{2}$, CH$_4$ and CO, which can be present in the gas, on small dust and on grown dust. Each of the volatile species $s$ is described by its surface density in the gas $\Sigma_{s}^{\rm gas}$, on small dust $\Sigma_{s}^{\rm sm}$, and on grown dust $\Sigma_{s}^{\rm gr}$.
At the post-processing, we also consider part of the ices on grown dust as the ices on pebbles $\Sigma_{s}^{\rm peb}$ (see Eq.~\eqref{eq:sigmaicepeb}). The distribution of volatiles in different phases over the disc can change due to the following three main processes: advection, collisional evolution of dust, and phase transitions. These processes are consecutively treated at every hydrodynamic step. First, the phase transitions are calculated as described below in this Section. Then, after the dust evolution step, where grain size is changing due to coagulation and fragmentation, ice mantles are redistributed between the dust populations in the same way as their refractory cores: the mass fraction of ice on small dust moved to the surface of grown dust equals to the mass fraction of small dust transformed into grown dust. At the third step, the volatiles are assumed to advect with their respective component (gas or dust), following the same Eqs.~\eqref{eq:1}--\eqref{energ} and~\eqref{contDsmall}--\eqref{momDlarge}, using the same third-order-accurate piece-wise parabolic method as for the gas and dust components \citep{2018A&A...614A..98V}.

The chemical processes in our modelling are treated as described in~\citet{Molyarova2021ApJ}. They include only adsorption of gas-phase species onto the surface of small and grown dust grains, and their thermal desorption, as well as photo-desorption by interstellar UV radiation. No gas-phase or surface reactions are included, and the only considered chemical processes are desorption and adsorption, which were shown
to be the most important for gas-phase abundances of most species \citep{2011MNRAS.417.2950I}. The chemical evolution of the surface densities of volatile species is calculated analytically from time-dependent balance equations ~\citep[Eqs.~(19)--(21) in][see also their Appendix~A]{Molyarova2021ApJ}:
\begin{eqnarray}
\frac{{\rm d}\Sigma_{s}^{\rm gas}}{{\rm d} t}&=&-\lambda_s\Sigma_{s}^{\rm gas}+\eta_s^{\rm sm}+\eta_s^{\rm gr},\label{eq:sig1}\\
\frac{{\rm d}\Sigma_{s}^{\rm sm}}{{\rm d} t}&=&\lambda_s^{\rm sm}\Sigma_{s}^{\rm gas}-\eta^s_{\rm sm},\label{eq:sig2}\\
\frac{{\rm d}\Sigma_{s}^{\rm gr}}{{\rm d} t}&=&\lambda_s^{\rm gr}\Sigma_{s}^{\rm gas}-\eta_s^{\rm gr},\label{eq:sig3}
\end{eqnarray}
where the mass rate coefficients per disc unit area of adsorption $\lambda_{s}$ and desorption $\eta_{s}$ for the species $s$ are calculated for local conditions at every hydrodynamic step, separately for small and grown dust populations. 
For convenience, we will further omit the indices referring to the species ($s$) and the dust population (``sm'' or ``gr'').

For each dust population, the desorption rate is a sum of thermal desorption and photo-desorption $\eta=\eta_{\rm td}+\eta_{\rm pd}$. Thermal desorption rate is calculated from the following expressions 
\begin{equation}
\label{eq:desorption_rate}
\eta_{\rm td} =  \widetilde{\sigma}_{\rm tot} N_{\rm ss} \mu_{\rm sp} m_{\rm p} k_{{\rm td}},
\end{equation}
\begin{equation}
\label{eq:desorption_rate_k}
k_{{\rm td}}=\sqrt{\frac{2 N_{\rm ss} E_{\rm b}}{\pi^2 \mu_{\rm sp} m_{\rm p}}} \exp{\left(-\frac{E_{\rm b}}{k_{\rm B}T_{\rm mp}}\right)},
\end{equation}
where $N_{\rm ss}=10^{15}$\,cm$^{-2}$ is the surface density of binding sites~\citep[characteristic value for amorphous water ice from][]{2017SSRv..212....1C}, $E_{\rm b}$ is the binding energy of the species to the surface, $\mu_{\rm sp}$ is the species molecular mass, $m_{\rm p}$ is mass of a proton.

The pre-exponential factor in Eq.~\eqref{eq:desorption_rate_k} is adopted as in \citet{1987ASSL..134..397T} and \citet{1993MNRAS.263..589H}, and is applied under the assumption of zeroth-order desorption, which seems to better describe desorption of CO and H$_2$O in TPD experiments~\citep{2001MNRAS.327.1165F,2005ApJ...621L..33O,2006A&A...449.1297B}. However, as was pointed out by~\citet{2022ESC.....6..597M}, this approach to calculation pre-exponential factors does not account for the rotational part of the partition functions of desorbing molecules and consequently underestimates the pre-exponential factor by a few orders of magnitude, especially for bigger molecules. This change in parameter value can lead to the positions of the snowlines shifted to colder regions by several astronomical units. However, the contribution of binding energy to the value of $k_{\rm td}$ is more significant, as it is included exponentially. We choose to keep the standard approach which was also adopted previously in~\citet{Molyarova2021ApJ}, and note that it affects the precise positions of the snowlines. The adopted values of binding energies are based on the experimental data for sublimation from icy surface \citep{2017SSRv..212....1C}. These are 5770\,K for H$_2$O, 2360\,K for CO$_{2}$, 1100\,K for CH$_4$ and 850\,K for CO.

To calculate the desorption rates from each dust population, we also need $\widetilde{\sigma}_{\rm tot}$ (cm$^2$\,cm$^{-2}$), which is the total surface area of dust grains (small or grown) per disc unit area. For our power-law size distribution ($p=3.5$) they are:
\begin{eqnarray}
\label{eq:surface}
\widetilde{\sigma}_{\rm tot}^{\rm sm}&=&\frac{3 \Sigma_{\rm d, sm}}{\rho_{\rm s}\sqrt{a_{\rm min} a_*}},\\
\label{eq:surface1}
\widetilde{\sigma}_{\rm tot}^{\rm gr}&=&\frac{3 \Sigma_{\rm d, gr}}{\rho_{\rm s}\sqrt{a_* a_{\rm max}}}.
\end{eqnarray}

Photodesorption rate by interstellar irradiation can be important in the outer disc regions with lower optical depth. We do not include the photo-desorption by stellar or accretion irradiation, because they are more efficient at  the inner radii and in the  disc upper layers, which  are not properly treated in our thin-disc approximation. We assume that the UV radiation field produced by the  star and the accretion region is zero in the midplane, where we consider our species to be, due to high optical depth. The photo-desorption rate 
can be estimated as
\begin{equation}
\label{eq:photodes}
    \eta_{\rm pd} =\widetilde{\sigma}_{\rm tot} \mu_{\rm sp}m_{\rm p}Y F_{\rm UV}.
\end{equation}
where $F_{\rm UV}$ (photons~cm$^{-2}$\,s$^{-1}$) is the UV photon flux and $Y$ (mol~photon$^{-1}$) is the photodesorption yield. We assume that the photodesorption yield is equal to one of water $Y = 3.5 \times 10^{-3} + 0.13  \exp\left({-336/T_{\rm mp}}\right)$ \citep{1995Natur.373..405W}.

The radiation flux $F_{\rm UV}$ is calculated for each disc point and is expressed in the units of standard UV field, $F_{\rm UV}=F_0^{\rm UV}G$.
The intensity of the interstellar UV radiation field corresponding to $G=1$ is $F^{\rm UV}_0=4.63\times10^7$\,photon~cm$^{-2}$\,s$^{-1}$~\citep{1978ApJS...36..595D}. For a disc situated in a star-forming region, a slightly elevated unattenuated UV field is applicable, so we assume $G_{\rm env}=5.5G_0$. For the disc midplane, which is illuminated from above and below, this field scales with the UV optical depth $\tau_{\rm UV}$ towards the disc midplane as
\begin{equation}
G_{\rm UV}=0.5 G_{\rm env} e^{-\tau_{\rm UV}},
\label{eq:gfactoruv}
\end{equation}
\begin{equation}
\tau_{\rm UV}= 0.5 \left(\varkappa_{\rm sm} \Sigma_{\rm d,sm} + \varkappa_{\rm gr} \Sigma_{\rm d,gr}\right).
\label{eq:tauUV}
\end{equation}
Here $\varkappa_{\rm sm}=10^4$\,cm$^2$\,g$^{-1}$, $\varkappa_{\rm gr}=2\times 10^2$\,cm$^2$\,g$^{-1}$ are typical values of absorption coefficients in the UV for small and grown grains~\citep[Fig.~1]{2019MNRAS.486.3907P}.

The adsorption rate depends on a total surface area of dust per unit volume $\sigma_{\rm tot}$, which is related to the total surface area of dust per unit disc surface $\widetilde{\sigma}_{\rm tot}$ through the scale height as $\sigma_{\rm tot}=\widetilde{\sigma}_{\rm tot}/2H$ \citep{1990MNRAS.244..432B}: Desorption rate formula requires the total surface area of grains, while for the adsorption rate
the total collisional cross section is used \citep{1990MNRAS.244..432B}, thus an additional factor of $1/4$ emerges and
\begin{equation}
\lambda= \frac{\widetilde{\sigma}_{\rm tot}}{8H} \sqrt{\frac{8k_{\rm B}T}{\pi \mu_{\rm sp} m_{\rm p}}}.
\label{eq:lambda}
\end{equation}

For more detailed description of the derivation of rate  coefficient of  adsorption and desorption, please see Section~2.3 in~\citet{Molyarova2021ApJ}.

We assume that initially in a prestellar core, small dust grains are covered with ice mantles, grown dust has not yet formed, and the volatiles are absent in the gas phase. The initial relative abundances of the volatiles are based on observations of ices in protostellar cores around the low-mass protostars \citet{Oberg2011a} and are related as H$_2$O : CO$_{2}$ : CO : CH$_4=100:29:29:5$. We choose to concentrate on considering main carbon- and oxygen-bearing species, so NH$_3$, which has the abundance comparable to that of CH$_4$, is not included. The initial mass fractions of ices relative to the refractory grain cores is $\approx 8.5\%$.

Fragmentation velocity $v_{\rm frag}$ is the key parameter that determines $a_{\rm frag}$, and hence the maximum dust grain size $a_{\rm max}$. To account for the effect of ice mantles on dust evolution, we adopt the fragmentation velocity $v_{\rm frag}$ depending on the presence of ice mantles on grown dust. At each time step, the values of $v_{\rm frag}$ are updated after calculating the surface densities of the volatiles. In collisions with velocities $<v_{\rm frag}$, grown dust grains coagulate, while in collisions with higher velocities, they fragment. In this paper, we assume that ice mantles  increase $v_{\rm frag}$ regardless of their composition~\cite[see][]{Molyarova2021ApJ}. This assumption is based on laboratory studies by~\citet{2009ApJ...702.1490W, 2015ApJ...798...34G}. Icy dust grains have $v_{\rm frag}=5$\,m~s$^{-1}$, and bare dust grains have $v_{\rm frag}=0.5$\,m~s$^{-1}$. Grown dust at a given location in the disc is considered icy if at that location, the total amount of ice on the grown dust is sufficient to cover all grown grains with at least one monolayer of ice. The thickness of the monolayer is estimated as the size of a water molecule $3 \times 10^{-8}$\,cm. 

\subsection{The considered models and their main parameters}

The simulation starts from a molecular cloud core with the initial profiles of gas surface density and angular velocity $\Omega$ describing typical prestellar cores 
\begin{equation}
\Sigma_{\rm g}(r)=\frac{r_{0}\Sigma_0}{\sqrt{r^2+r_0^2}},
\end{equation}
\begin{equation}
\Omega=2\Omega_0 \left(\frac{r_0}{r}\right)^2\left(\sqrt{1+\left(\frac{r}{r_0}\right)^2}-1\right).
\end{equation}

Initially, all refractory material is in the form of small dust, with the surface density calculated assuming dust-to-gas mass ratio of 0.01. The calculation was performed on a $N_r \times N_{\phi} = 390 \times 256$ polar grid with the outer boundary of the computational area $R_{\rm out}$, and the inner boundary  $R_{\rm in}$. The central disc region inside $R_{\rm in}\approx0.63$\,au is replaced by a sink cell \citep[see][]{2018A&A...614A..98V}. We assume that about 10\% of the matter accreted to the protostar is ejected as jets and outflows. By the end of the simulation, a small amount of mass remains in the envelope. The main parameters of the considered models are summarised in Table~\ref{tab:model}.

We consider two models with different initial cloud masses: $\rm M_{\rm core} =$ 0.66\,$M_{\odot}$ and 1\,$M_{\odot}$. We use M1 model ($\rm M_{\rm core} =$ 0.66\,$M_{\odot}$) as the reference model, and M2 model to investigate how the initial cloud mass affects our conclusions (see Section~\ref{sec:mass}). The disc mass relative to the stellar mass at the end of the simulations is quite high and approximately the same in the two models (0.55 and 0.60, respectively), i.e., both discs are gravitationally unstable, but the total mass of the system also affects gas and dust evolution. The disc is formed at 53\,kyr in M1 model and at 79\,kyr in M2 model.

\begin{table*}
\caption{Model parameters. $M_{\rm core}$ is the initial mass of the core, $\beta$ is the ratio of rotational energy to gravitational energy, $T_{\rm init}$ is the initial temperature of the gas equal to the temperature of the background radiation $T_{\rm bg}$, $\Omega_0$ is the characteristic angular velocity of rotation of the core, $\Sigma_0$ is the gas surface density at the centre of the core, $r_0$ is the radius of the central plateau in the initial core, $R_{\rm out}$ is the outer boundary of the computational area. $M_{\star}$ and $M_{\rm disc}$ are the masses of the central star and disc at the end of the simulation (500\,kyr).}
\label{tab:parameters}
 \begin{tabular}{lccccccccc}
 \hline\noalign{\smallskip}
Model & $M_{\rm core}$ & $\beta$ & $T_{\rm init}$ & $\Omega_0$ & $\Sigma_0$ & $r_0$ & $R_{\rm out}$ & $M_{\rm star}$ (0.5 Myr) & $M_{\rm disc}$ (0.5 Myr) \\
& ($M_{\odot}$) & (\%) & (K) & (km s$^{-1}$ pc$^{-1}$) & (g cm$^{-2}$) & (au) & (au) & ($M_{\odot}$) & ($M_{\odot}$) \\
 \hline\noalign{\smallskip}
M1 & 0.66 & 0.28 & 15 & 2.26 & $0.173$ & 1029 & 6116 & 0.40 & 0.22 \\
M2 & 1.00 & 0.28 & 15 & 1.51 & $0.115$ & 1543 & 9170 & 0.58 & 0.35 \\
 \noalign{\smallskip}\hline
 \label{tab:model}
\end{tabular}
\end{table*}

\section{RESULTS}
\label{sec:res}
\subsection{Global disc evolution}

\begin{figure*}
\center\includegraphics[width = 2\columnwidth]{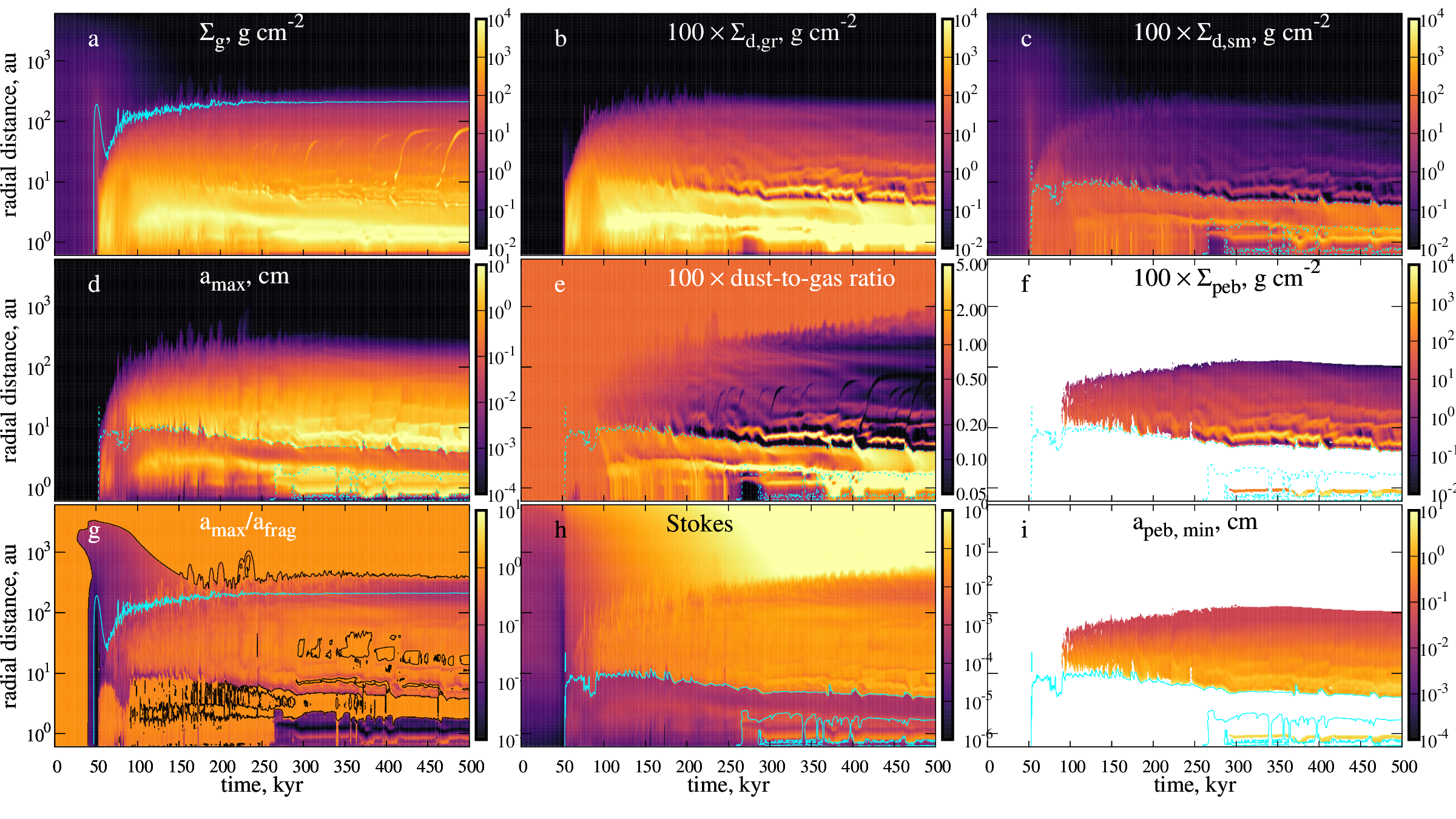} 
\caption{Evolution of the azimuthally averaged radial distributions of gas and dust parameters. The panels show (a) surface density of gas, (b) grown dust (including pebbles), (c) small dust, (d) maximum dust size, (e) dust-to-gas mass ratio, (f) surface density of pebbles, (g) maximum dust size relative to the fragmentation barrier, (h) Stokes number, (i) minimum pebble size. The solid blue line in panels (a) and (g) indicates the approximate disc boundary at $\Sigma_{\rm g} = 1$~g~cm$^{\rm -2}$, the dashed cyan line in panels (c)--(f), (h), and (i) shows the position of the water snowline. The black contour in panel (g) corresponds to $a_{\rm max}/a_{\rm frag} = 0.9$ and roughly delineates the regions where dust size is restricted mainly by fragmentation.}
\label{ris:2}
\end{figure*}

Figure~\ref{ris:2} shows the global evolution of the disc main parameters in M1 model. The gas and dust surface densities evolve forming a system of ring-like structures in the disc at $150-200$\,kyr, seen as bright yellow horizontal lines in panels (a), (b), and (c). The dust size changes dramatically at the water snowline indicated by the cyan dashed line (see  panel (d) of Figure~\ref{ris:2}). Outside the snowline, maximum dust size ($a_{\rm max}$) reaches centimetres, while inside the snowline it is only fractions of millimetre. In addition, at $\approx 250$\,kyr, an about 1\,au wide dust ring is formed at approximately 1\,au distance.
It appears as a region of higher gas and dust surface densities in the three upper panels of Figure~\ref{ris:2}, with visible changes in dust size, $a_{\rm max}/a_{\rm frag}$ and Stokes number at the same times ($220-500$\,kyr) and radial distances ($\approx0.8-2$\,au). Inside this ring, water is in the ice phase (note that the specified region is circumvented by a dashed blue line marking the snowline position) and dust size also reaches centimetres. In panel (g) we can see that $a_{\rm max}$ inside this ring is significantly lower than $a_{\rm frag}$, as well as in the regions outside the water snowline. This is due to the fact that in the inner ice ring around 1\,au and in almost all regions beyond the water snowline collisional dust fragmentation is less efficient and dust evolution is dominated by radial drift. These regions  turn out to be most favourable for the formation of pebbles, as we see in panels (f) and (i). In this context, the ring at $\sim1$\,au is particularly noteworthy, suggesting possible connection with the formation of the terrestrial planets of the Solar System, which have  similar radial distances.

In some disc regions dust becomes large enough to be defined as pebbles (see Section~\ref{sec:2B}). Panel (f) of Figure~\ref{ris:2} shows that pebbles appear at $\rm t \approx 90$\,kyr at distances of about 10 to 50\,au. This region begins to expand and after 300\,kyr is located between 5 and $\approx 100$\,au. This region is also characterised by the formation of dust rings. In the ring at $\approx10$\,au shifting with time parallel to the water snowline, the surface density of pebbles $\Sigma_{\rm peb}$ reaches $10^{1}$\,g~cm$^{-2}$. In addition, after $\approx 300$\,kyr, a narrow ring of pebbles forms at 1\,au at the inner edge of the inner ice ring. As can be seen from the position of the water snowline in panels (f) and (i) of Figure~\ref{ris:2}, pebbles in the disc exist only in regions where water is in the ice phase.

\begin{figure*}
\center\includegraphics[width = 2\columnwidth]{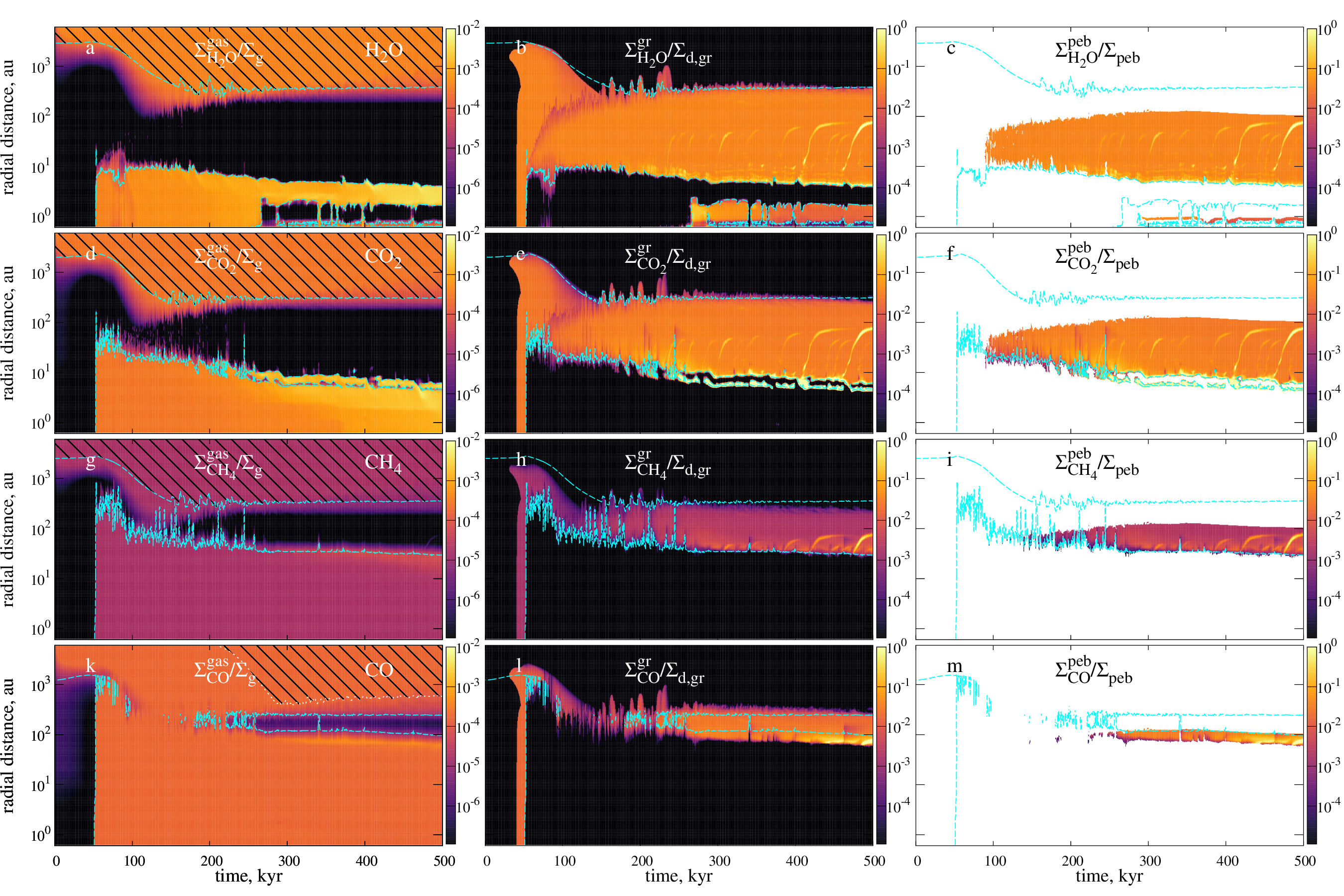} 
\caption{Evolution of the azimuthally averaged radial distributions of the four volatiles in the gas and in the ice on grown dust and on pebbles. For each panel, the dashed cyan line shows the equilibrium position of the snowline. The black shading marks the regions where photo-dissociation by interstellar UV radiation should be efficient enough to destroy the gas-phase species \citep[see Section~3.3.1 in][]{Molyarova2021ApJ}.}
\label{ris:1}
\end{figure*}

Figure~\ref{ris:1} shows the evolution of the azimuthally averaged radial distributions of four volatiles (H$_2$O, CO$_{2}$, CH$_{4}$ and CO) in the gas, on grown dust and on pebbles. The panels on the left show the fraction of these volatiles in the gas. After $250-300$\,kyr their abundance is elevated in the inner regions of the disc, compared to the earlier stages (around 100\,kyr). The abundances of gas-phase H$_2$O and CO$_{2}$ inside their snowlines increase gradually and reach values an order of magnitude higher than immediately after the disc formation. These additional volatiles are brought from the outer disc regions by the drift of grown dust grains \citep{2012A&A...539A.148B}, bringing ice mantles that subsequently sublimate into the gas phase \citep{2017A&A...608A..92D,2018ApJ...864...78K}. The snowlines of CH$_{4}$ and CO are further from the star due to their lower desorption energies. At these larger distances, dust grains are smaller and have lower surface density (see panels (b) and (d) of Figure~\ref{ris:2}), so the effect of the drift bringing in the volatiles is weaker. However, the fraction of these species in the gas is also elevated at the inner side of their snowlines, as was previously shown by \citet{Molyarova2021ApJ}.

The central panels of Figure~\ref{ris:1} show the distribution of volatiles on grown dust. In most of the disc, their mass fractions are close to the initial values for the ices on small dust at the onset of the collapse. However, near the snowlines, there are regions with the increased amount of ice on grown dust. For H$_2$O and CO$_2$ this effect is more pronounced, and their surface density at some time instances exceed that of the grown dust. These snowlines are situated at smaller distances, hence more volatiles that are brought with drifting dust can accumulate at the snowline. These are also the species that have multiple snowlines. At the radial distances where these species tend to freeze (inside 10\,au), the radial distribution of the temperature is substantially non-monotonous due to the presence of substructures, particularly, of the dense dust rings.
Inside the dust rings, the densities are higher and temperatures are lower compared to the surrounding regions \citep[see, e.g.,][]{Kadam2022MNRAS}. This creates the conditions favourable for the freeze-out of volatiles inside the rings, which occurs for H$_2$O at $\approx1$\,au, and for CO$_2$ at $\approx6$\,au.

The right panels of Figure~\ref{ris:1} show the distribution of volatiles on pebbles. We see that H$_2$O is present on pebbles in both inner and outer regions where pebbles exist. CO$_2$ ice is present on pebbles in almost all the  regions where pebbles exist as well, however the inner boundary of the CO$_2$ ice spatial distribution on pebbles begins a few astronomical units farther from the star. Between the H$_2$O and CO$_2$ snowlines, only H$_2$O is present on pebbles, as well as in the inner ring at 1\,au. At early times between $\approx100-220$\,kyr, small amount of CH$_4$ ice appears on pebbles, after that it is constantly present at the distances from about 30 to 80\,au. CO ice on  pebbles appears at later times, after 250\,kyr, at distances from about 80 to 120\,au, and by 500\,kyr this region shifts to $60-100$\,au. At some radii, the surface density of CO on pebbles exceeds the surface density of pebbles themselves. 

The adsorption and desorption processes have finite non-zero time scales, so the current balance of volatile content in the gas and in the ice does not necessarily correspond to the equilibrium solution. This can be seen in Figure~\ref{ris:1}, where the position of the equilibrium snowlines, calculated under the simplifying assumption of equal desorption and adsorption rates, is shown by the cyan lines. The equilibrium snowline position are calculated 1) for azimuthally averaged distributions and 2) assuming equilibrium between adsorption and desorption rates \citep[see Section 2.4 in][]{Molyarova2021ApJ}. In some cases, ice is present beyond the region delineated by the equilibrium snowline in Figure~\ref{ris:1}, namely inside the snowline. In particular, CO ice is present $\approx10$\,au closer to the star than it is expected from its equilibrium snowline defined by thermal desorption. This indicates a significant contribution of non-steady-state and non-axisymmetric effects to the distribution of volatiles, especially in the outer regions of the disc. The main dynamical factors leading to the deviation from equilibrium are the presence of spirals and the radial drift of icy dust grains.

The development of the magnetorotational instability~\citep[MRI,][]{Balbus1991ApJ,1998RvMP...70....1B,Turner2014prpl} in the inner regions of the disc can result in the occurrence of accretion and luminosity bursts~\citep{2010ApJ...713.1134Z,2020ApJ...895...41K,2020A&A...644A..74V}. Such bursts are caused by the increase in turbulence and are taken into account by the associated increase in $\alpha_{\rm eff}$ in our models. Accretion and luminosity bursts self-consistently appear in our simulations for as long as the conditions for the MRI development are fulfilled (see Section~\ref{sec:visc_model}). Accretion outbursts with tens of $L_{\odot}$ magnitude occur regularly during the first 250\,kyr of disc evolution. The luminosity outburst accompanying the growth in the accretion rate heats the disc and leads to thermal desorption of ices, temporarily shifting the snowlines of all volatiles farther away from the star \citep{2016Natur.535..258C,2013A&A...557A..35V,2023MNRAS.521.5826H}. Such shifts can be seen as ``spikes'' in the positions of snowlines in Figure~\ref{ris:1}. Bright outbursts inevitably affect the composition of ice mantles on dust grains, including pebbles. A more detailed investigation of the effect of the outbursts on the composition of ices is beyond the scope of this paper and merits a separate study.

\subsection{Composition of solids}
\label{sec:solid_composition}

The solid matter in the disc consists of the following components: small dust, grown dust and pebbles, and ices (H$_{2}$O, CO$_{2}$, CH$_{4}$, and CO) covering all these types of dust grains.  We will refer to small dust, grown dust and pebbles as refractory components, meaning that in our model, they are in the solid phase at any conditions in the disc, unlike the volatiles that can be either in the ice or in the gas. In our modelling, pebbles are part of grown dust population. In order to distinguish pebbles from grown dust of smaller sizes, we will hereafter refer to the mass or mass fraction of grown dust meaning that of grown dust excluding the pebbles, unless otherwise stated.

\begin{figure*}
\center\includegraphics[width = 2\columnwidth]{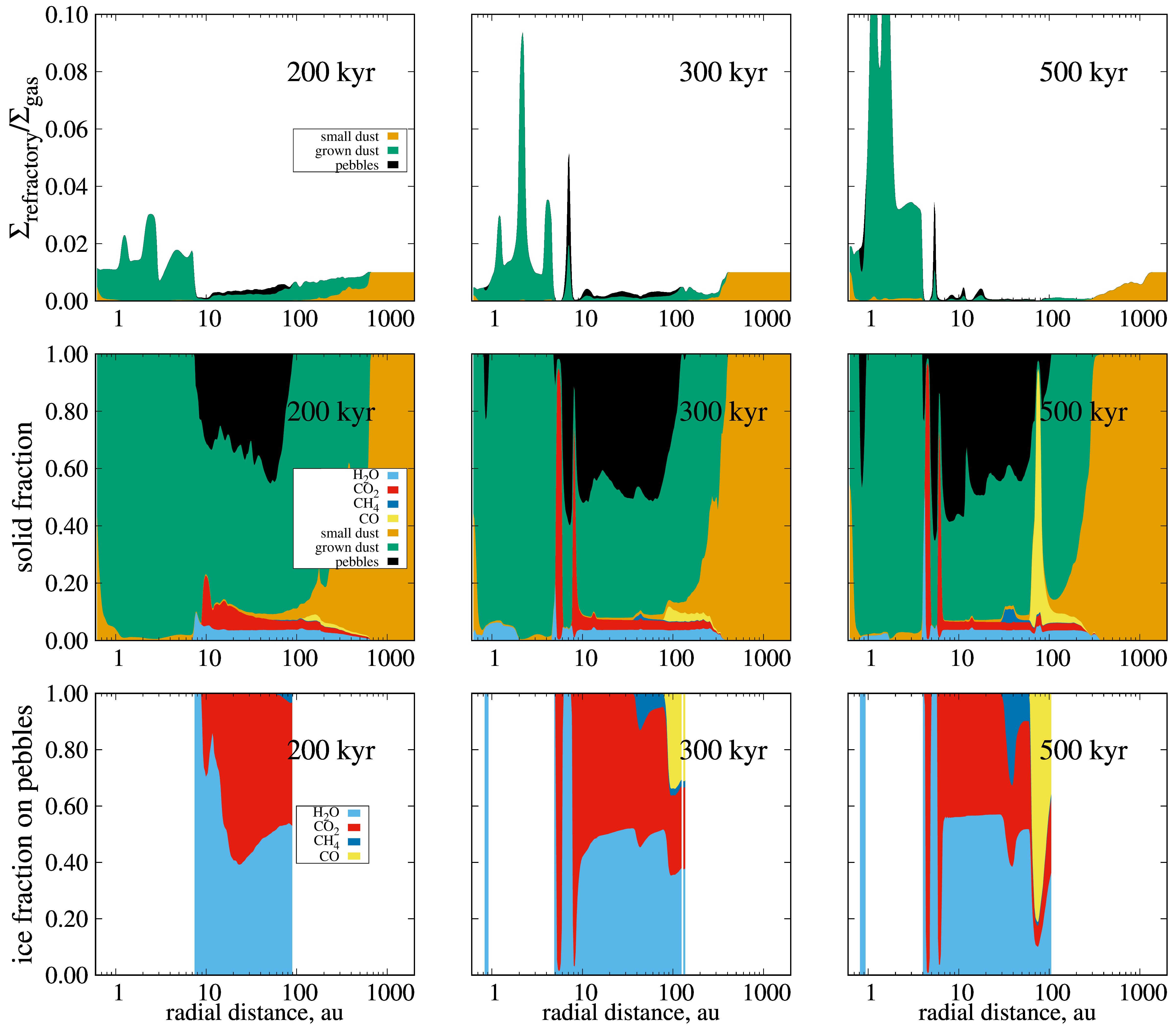}
\caption{The top three panels show the mass fraction of the refractory component (small dust, grown dust, pebbles) relative to the gas. The middle row shows the fraction of ice and refractory components in total amount of solids. The lower three panels show the fraction of ice in the composition of ice mantles on pebbles.}
\label{ris:3}
\end{figure*}

The top panels of Figure~\ref{ris:3} show the cumulative mass fraction of refractory components relative to gas as a function of distance from the star. It shows three characteristic time instances: an early gravitationally unstable disc (200\,kyr); an intermediate stage of disc evolution (300\,kyr); and an evolved nearly axisymmetric disc (500\,kyr). The mass fraction of refractories relative to gas evolves from the initial value of 0.01 as a result of dust growth and dynamics in the disc. The initial value is only kept in the surrounding envelope beyond $\sim1000$\,au, where collisions are rare and dust grains do not deviate much from their initial state in the cloud. In the inner $\approx$10\,au, the fraction of refractories becomes a few times higher, while in the disc outside this distance the refractory component is depleted. This distribution is a result of the radial drift of grown dust combined with the variable $v_{\rm frag}$ in our model. Dust drift efficiently brings the solid material to the inner disc \citep{1977MNRAS.180...57W}, but its rate depends on grain size and fragmentation hampers this process~\citep{2012A&A...539A.148B}. The change of $v_{\rm frag}$ at the water snowline ($5-10$\,au depending on time) decreases $a_{\rm frag}$ and the drift slows down, leading to accumulation of grown dust in this inner region.

At distances from the water snowline to $\approx 100$\,au, a noticeable fraction of solid matter is in pebbles. These are the distances where grown dust can reach larger sizes due to higher $v_{\rm frag}$ of ice-covered grains. Their mass fraction relative to the gas is mostly below 0.005, however inside the rings rich in dust, e.g. at 8\,au after 300\,kyr, or at 1\,au at 500\,kyr it can exceed 0.01. The results of~\citet{Vorobyov2022} show similar amount of pebbles for $\alpha$-parameter values $\alpha=10^{-3}$ and $10^{-4}$, which are close to the values in our modelling. However, for $\alpha=10^{-4}$ their modelling also produces a high fraction of pebbles in the inner disc regions, which in our results are absent, although inside 10\,au is where the dead zone is situated, with values of $\alpha$ as low as $\sim10^{-5}$. This is the result of $v_{\rm frag}=3$\,m~s$^{-1}$ for all dust grains in~\citet{Vorobyov2022}, which is higher than our value for bare grains ($0.5$\,m~s$^{-1}$), thus providing less fragile dust in the inner disc regions in their modelling. We conclude that pebbles constitute a significant fraction of solid material in most of the disc, and their existence is favoured by the presence of ice mantles.

Volatiles are also an important component of the solid material. In the regions where dust grains are covered with mantles, these mantles typically consist of hundreds of monolayers of ice. The middle row of Figure~\ref{ris:3} shows the fraction of ice and refractory components in the solid phase across the disc. At the selected times, CO$_2$ and H$_2$O dominate the ice composition in most of the disc. There are peaks in the species ice abundances associated with their snowlines, where their mass fraction becomes significant in the total ice composition. By 500\,kyr, the main water snowline is at 4\,au, with a narrow peak where H$_2$O comprises half of the local solid mass. In two narrow annuli at $r\approx6-8$\,au at 300\,kyr and at $r\approx4-6$\,au at 500\,kyr, CO$_2$ dominates in the composition of solids. At late times and at distances of about 100\,au, there is an annulus where more than 90\% of the mass of solids is in CO ice. Ice accumulation at these radii is associated with the snowlines of the corresponding volatiles. Dust drifting relative to the gas brings volatiles from the outer regions of the disc, and the azimuthal variations in the radial velocity of gas and dust allow the volatiles to move back to the ice phase in the vicinity of the snowlines \citep[see Figure~5 and Section~3.3.1 in][]{Molyarova2021ApJ}. 

The most notable peaks of ices are presented by the two annuli of CO$_2$ ice. They surround a prominent dust ring rich in pebbles, which begins to form after 200\,kyr. These ice annuli appear due to the temperature contrast between the ring and the nearby regions creating multiple snowline structure. The temperature in the regions immediately inside and outside this ring is determined by the irradiation from the star and accretion, but inside the ring itself, the temperature is higher. Due to elevated gas surface density in the ring, which can be seen in Figure~\ref{ris:2}, the contribution of viscous heating is more significant, and radiative cooling is less efficient because of higher optical depth of dust. As a result, CO$_2$ is sublimated inside the dense ring and frozen in the adjoining regions. The mass fraction of CO$_2$ ice in the solid component around the dust ring is high due to lowered dust surface density outside the ring, while the absolute surface density of CO$_2$ ice is fairly the same. This means that the peaks of CO$_2$ appear in the middle row of Figure~\ref{ris:3} because of depletion of dust, which drifts towards the ring's pressure maximum, rather than because of the overabundance of CO$_2$ ice.

The lower panels of Figure~\ref{ris:3} show the relative ice fraction in ice mantles on pebbles. At 200\,kyr, the ice on pebbles consists mainly of H$_2$O and CO$_2$, with approximately equal mass fraction except for the specific ice-rich annuli described above. The pebbles formed at 1\,au are covered exclusively by H$_2$O ice. Throughout the disc lifetime, water is always present in the composition of ices on pebbles, making up about half the mass fraction of the ice mantle. At later times, other volatiles start to dominate the composition of ices on pebbles in the vicinity of their respective snowlines. As the disc cools down with time, the snowlines move closer to the star, and more volatile species CH$_{4}$ and CO are able to freeze-out and contribute to the composition of ice mantles on pebbles, which are only present inside $\approx100$\,au. By the end of the simulation, CO becomes the dominant species on pebbles beyond 60\,au, while H$_2$O and CO$_2$ remain the main ices on pebbles inside this distance.

\begin{figure*}
\includegraphics[width = \columnwidth]{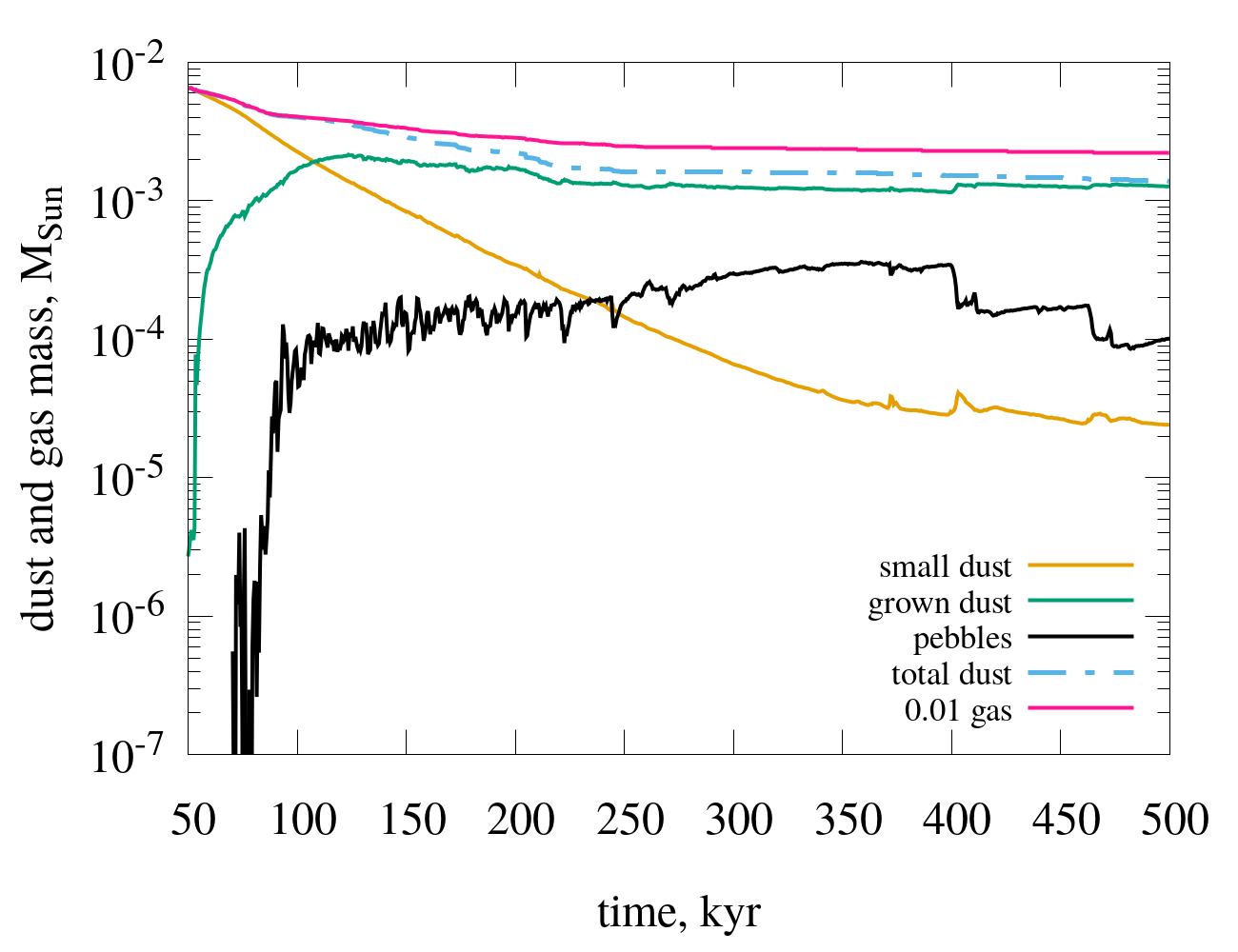}
\includegraphics[width = \columnwidth]{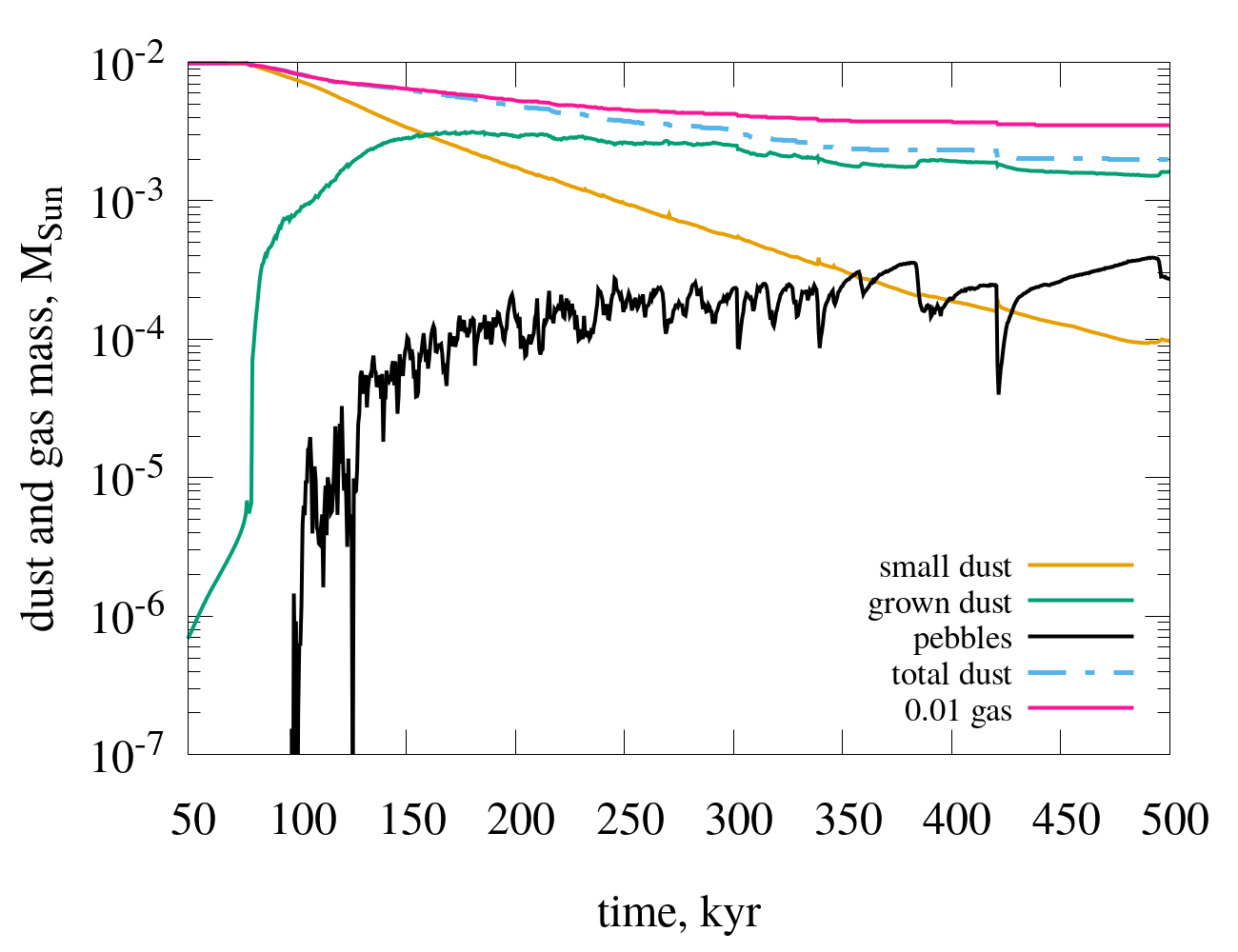}
\caption{Evolution of the model-integrated mass of small dust, grown dust and pebbles, total dust and gas. Left panel is for M1 model ($M_{\rm core} = 0.66 M_{\rm \odot}$), right panel is for M2 model ($M_{\rm core} = 1 M_{\rm \odot}$).}
\label{ris:5}
\end{figure*}

\begin{figure*}
\includegraphics[width = \columnwidth]{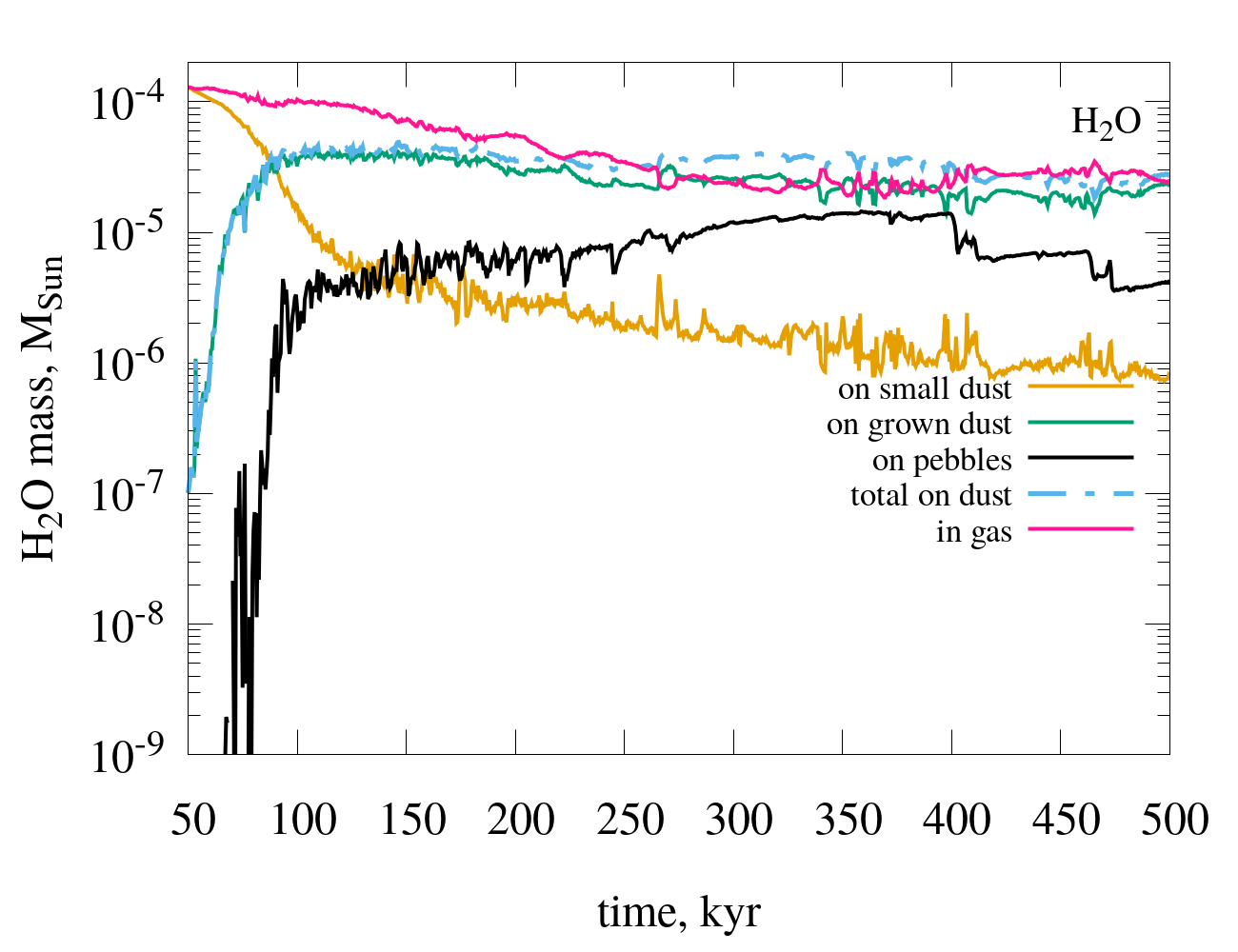}
\includegraphics[width = \columnwidth]{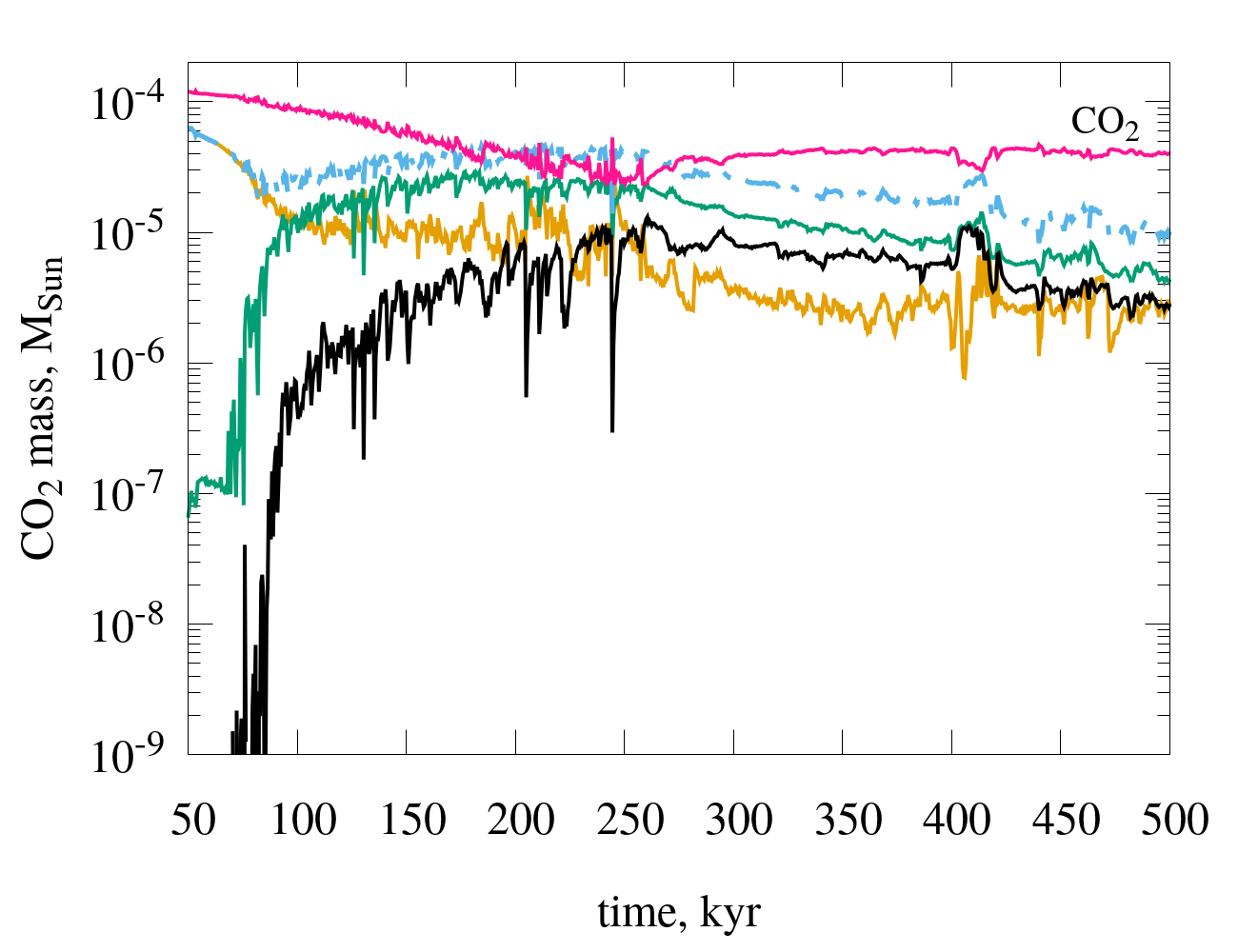} \\ 
\includegraphics[width = \columnwidth]{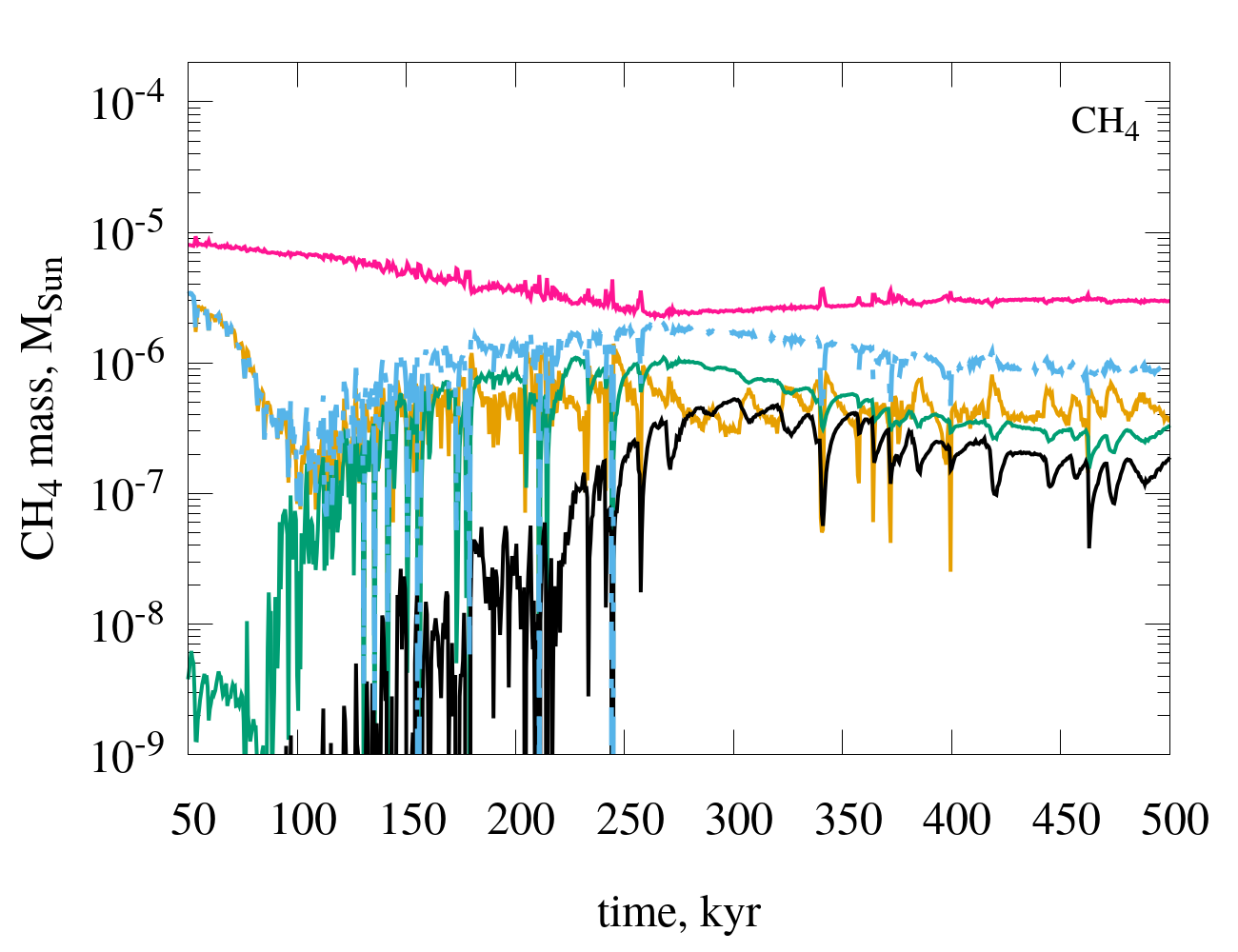}
\includegraphics[width = \columnwidth]{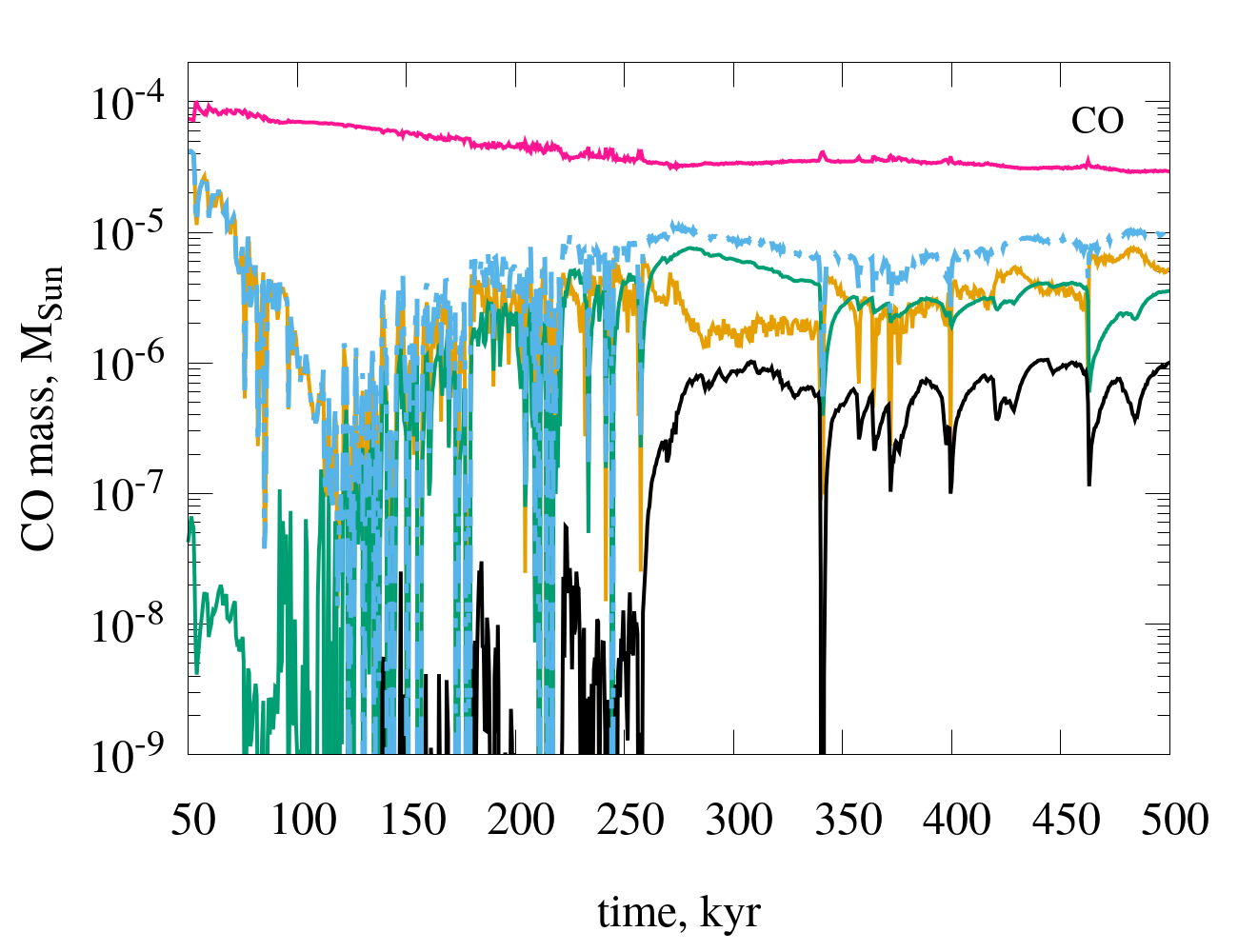}
\caption{Evolution of the model-integrated mass of volatiles (H$_{2}$O, CO$_{2}$, CH$_{4}$ and CO) on small dust, on grown dust, on pebbles, total in the solid phase and in the gas.}
\label{ris:6}
\end{figure*}

\begin{figure*} 
\includegraphics[width =0.65\columnwidth]{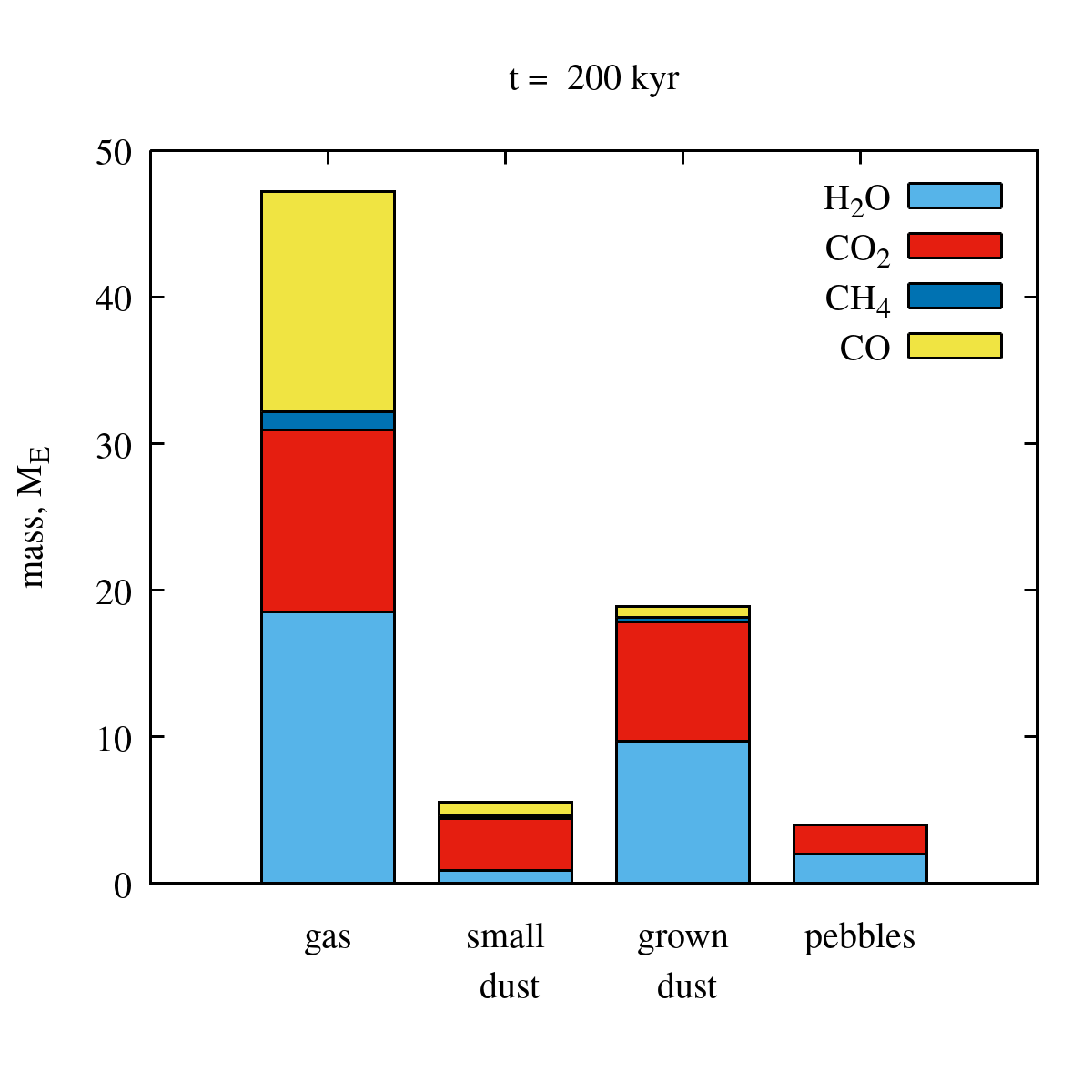}
\includegraphics[width =0.65\columnwidth]{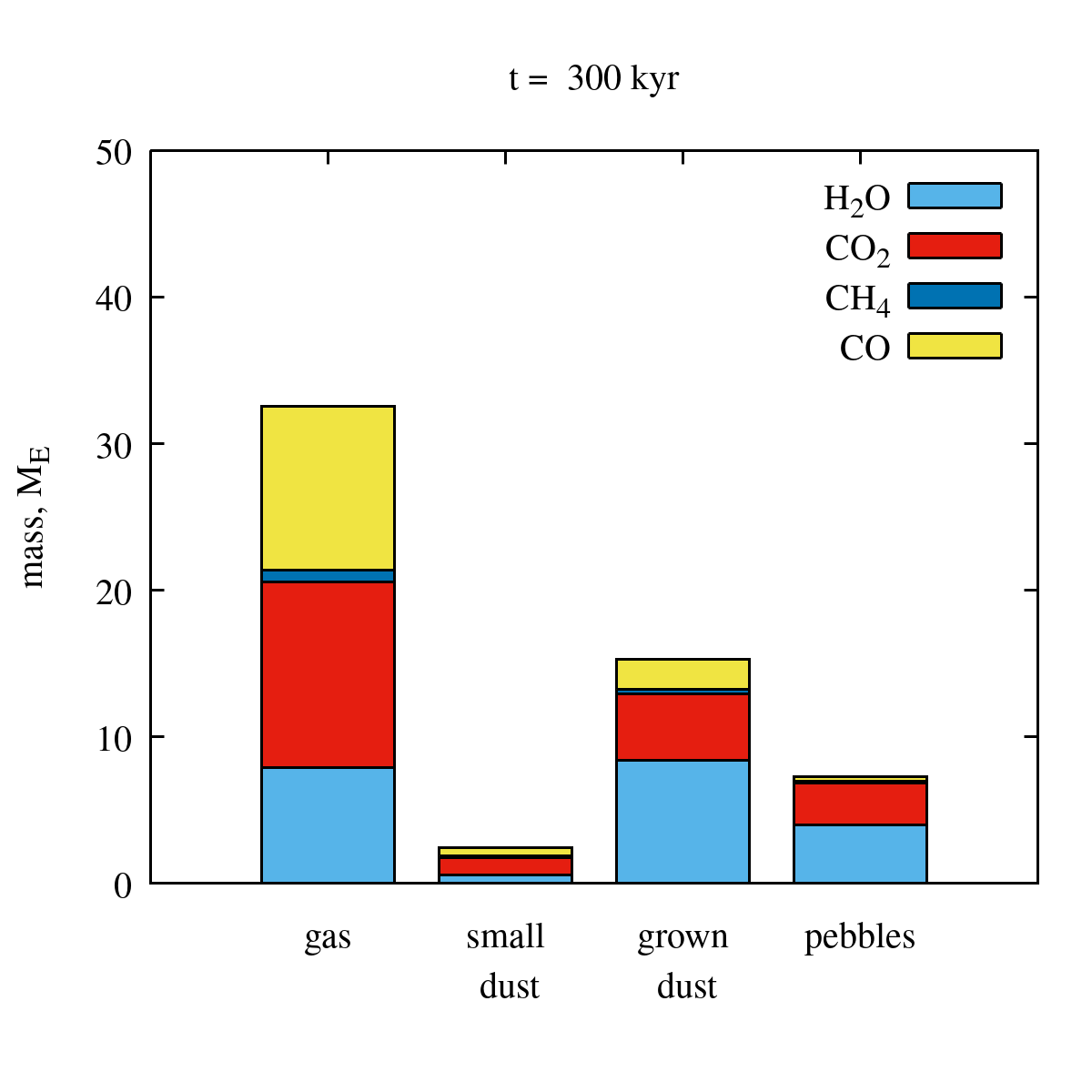}
\includegraphics[width =0.65\columnwidth]{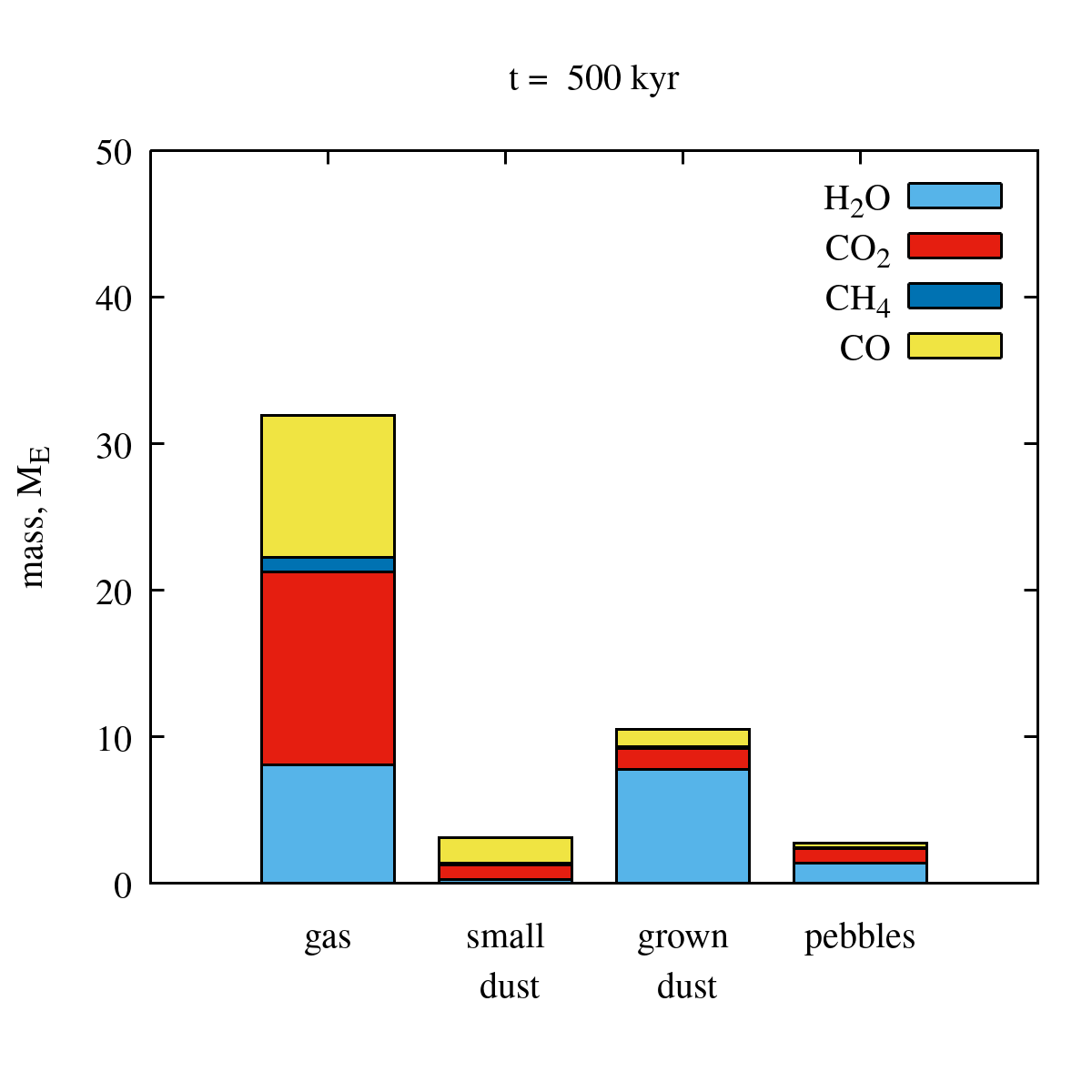}\\
\includegraphics[width =0.65\columnwidth]{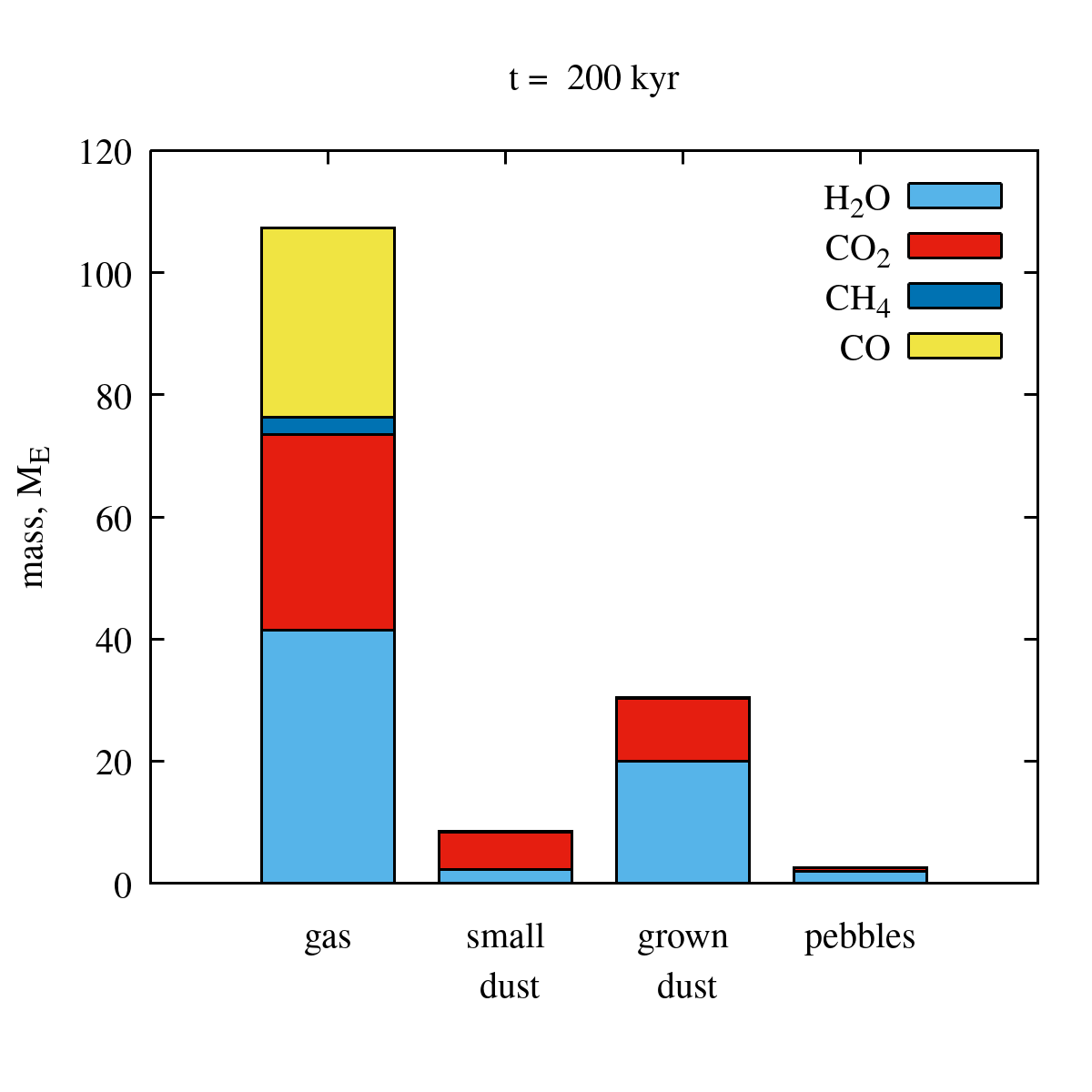}
\includegraphics[width =0.65\columnwidth]{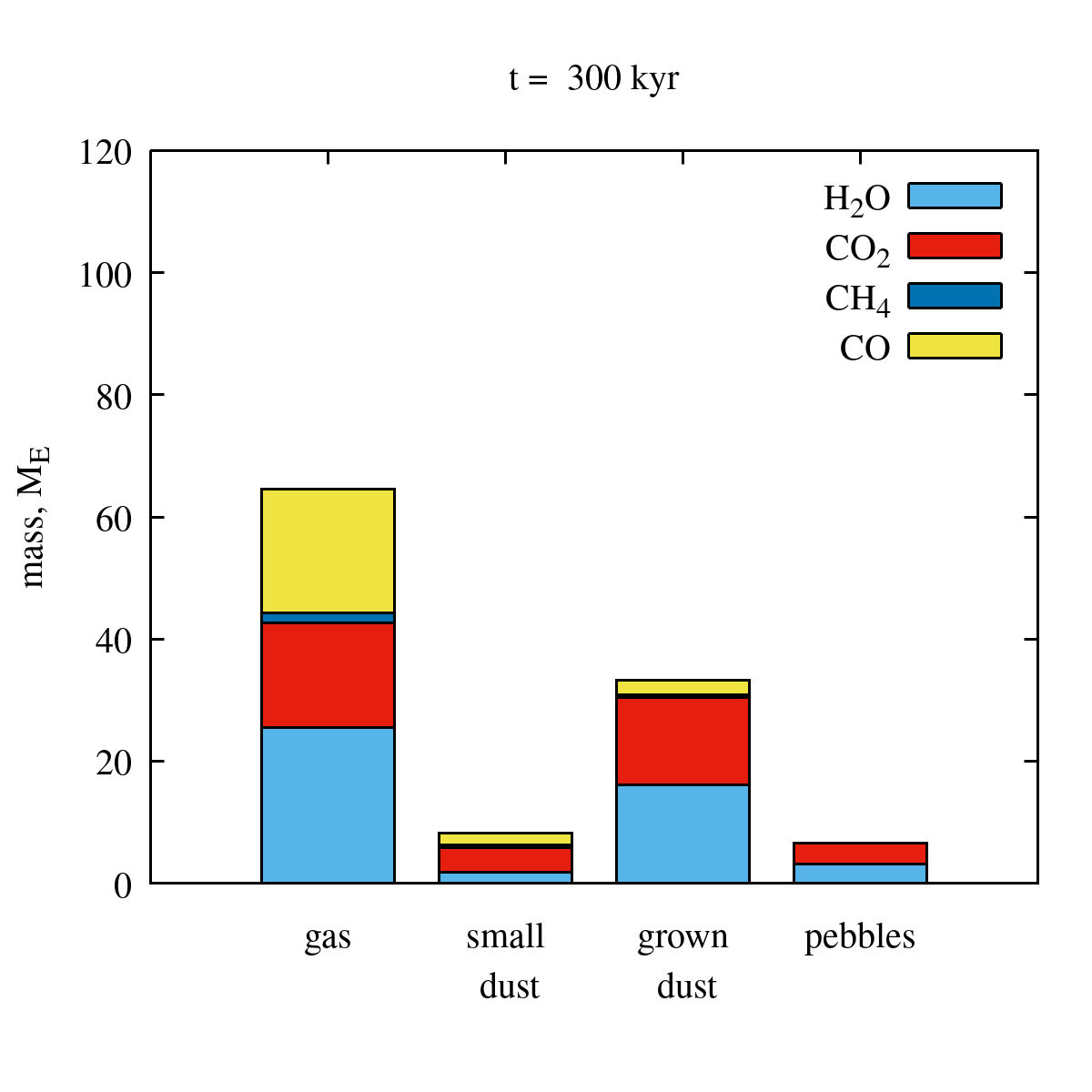}
\includegraphics[width =0.65\columnwidth]{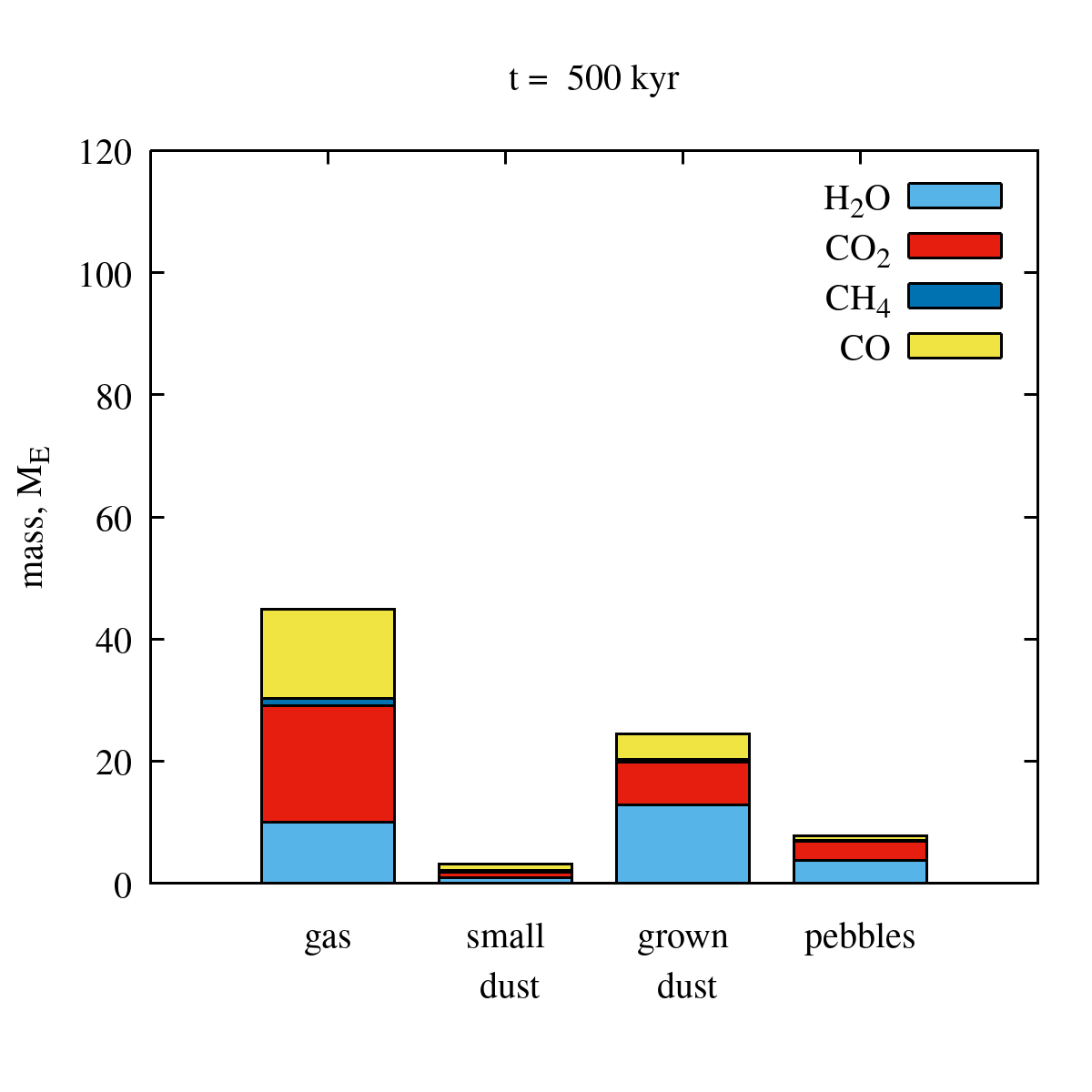}
\caption{The masses of volatiles in gas, on small dust, on grown dust, and on pebbles in $M_{\oplus}$ at 200, 300, and 500\,kyr. The top panel shows the results for M1 model, the bottom panel for M2 model. Note the different range on the $y$-axis for M1 and M2 models.}
\label{ris:ices_total}
\end{figure*}

Figure~\ref{ris:5} shows the evolution of the total mass of gas and various solid fractions. We integrated surface densities of the components over the entire computational area, which means that in addition to the disc, they include the matter from the outer envelope. Since the envelope mostly consists of primordial matter that hardly evolved from the initial state in terms of dust size, the total masses of pebbles and grown dust in the model are almost identical to those inside the disc. For small dust and gas this difference is more substantial, with a few per cent of mass belonging to the envelope. We note that the model includes accretion from the disc, which means that some matter, including gas, dust and volatiles leaves the disc and contributes to the content of the central sink cell, protostar, with a fraction of mass ejected with jets. Thus, the total masses of gas and dust in Figure~\ref{ris:5} are declining.

It can be seen that the mass of small dust gradually decreases, while the mass of grown dust first increases due to its formation by coagulation of small dust, reaching $2 \times 10^{-3} M_{\odot}$, then gradually decreases due to radial drift of grown grains to the star. As for pebbles, they are present during almost the entire disc lifetime. At 100\,kyr, only 50\,kyr after the disc formation, the pebble mass reaches a value of $10^{-4} M_{\odot}$ and increases slightly with time, but by the end of the simulation falls back to this value. The maximum pebble mass is reached at 400\,kyr and is about $3.5\times 10^{-4} M_{\odot}$ ($\approx115$\,$M_{\oplus}$), which is comparable to the mass of solids in the Minimum Mass Solar Nebula \citep[$30-300$ $M_{\oplus}$,][]{1977Ap&SS..51..153W}. The pebbles are abundant from the very early stages of disc formation, which means that the conditions for planet formation exist throughout the lifetime of the disc. However, whether these pebbles can immediately start forming planetesimals, e.g., through the streaming instability \citep{2005ApJ...620..459Y} remains an open question and requires a separate consideration.

As for M2 model, which has $M_{\rm core} = 1 M_{\rm \odot}$, the total mass of gas and dust in it is $1.5$ times higher than in M1 model described above. The integrated amount of pebbles formed in the models is only slightly different,  as shown in the right panel of Figure~\ref{ris:5}. It reaches $10^{-4} M_{\odot}$ later, by $\approx150$\,kyr, which is $\approx70$\,kyr after the disc formation. The maximum value of $3\times 10^{-4} M_{\odot}$ is reached by 490\,kyr. The masses of pebbles in the two models are of the same order, but the timing  is different, the more massive M2 model characterised by slower evolution.

Figure~\ref{ris:6} presents the evolution of the integrated mass of volatiles in different phases. 

Although pebbles are part of grown dust and locally they have identical relative fractions of ices, globally the compositions of their mantles are different because grown dust exists in a wider region of the disc. The integrated mass of water ice on grown dust is several times larger than that on pebbles, but this difference decreases with time, reaching its minimum at 400\,kyr. At 400\,kyr, as well as 460\,kyr, the amount of pebbles itself decreases sharply (see Figure~\ref{ris:5}), leading to simultaneous decrease of the mass of water ice on pebbles.

The masses of CO$_{2}$ and CH$_{4}$ ices on grown dust are not significantly different from their masses on pebbles after $250-300$\,kyr. The mass of CO ice on grown dust during the entire disc lifetime exceeds the mass of CO ice on pebbles by several orders of magnitude. Overall, ice mantles on grown dust are more rich in CH$_{4}$ and CO than ice mantles on pebbles.

While CO is almost absent on pebbles, in the gas its integrated mass is comparable to the masses of gas-phase water and CO$_{2}$. Since CO is more volatile, it is in the gas phase in most parts of the disc. The total abundance of carbon in the gas is higher than in the ice, suggesting a higher C/O ratio. Small dust mantles are dominated by CO$_{2}$ and CO ices, although the CO ice mass changes frequently due to outbursts and is more sensitive to them. Compared to CO$_{2}$ and CO, water and methane masses on small dust are lower by a factor of several. For water this is because small dust is dominant in the outer regions of the disc, while water ice is depleted there and brought closer to the star with inward-drifting grown dust grains. For methane the low mass is due to its low initial abundance. Most of the mass of the volatiles belongs in the gas phase, the only exception being H$_2$O, which mostly dominate in the ice phase after $\approx250$\,kyr.

The integrated masses of ices on pebbles compared to the integrated masses of volatiles in the gas, on small dust, and on grown dust at different time points are presented in Figure~\ref{ris:ices_total}. The total ice composition is dominated by oxygen-rich H$_{2}$O and CO$_{2}$. The contribution of CO is more significant for the ice on grown dust, and especially for the ice on small dust. This means that these ice mantles contain relatively more carbon than the ice mantles on pebbles. In the gas composition, one third of the mass of volatiles is encapsulated in CO, making the gas phase the most carbon-rich fraction, although oxygen still dominates the elemental composition. Thus, ices on pebbles demonstrate characteristically lowered carbon abundance compared to ices on grown and small dust grains, as well as the volatiles in the gas.

In M2 model, the total mass of ice mantles on pebbles is of the same order as in M1 model, if not slightly lower (see lower  panels in Figure~\ref{ris:ices_total}). Although total mass of volatiles is higher in M2 model due to higher initial core mass, the pebble masses have similar values (see Figure~\ref{ris:5}). The mass fraction of ices on pebbles are also similar between the models, leading to the close values of total ice masses on pebbles. An exception is presented by the 500\,kyr time instance, when in M2 model the ice mass on pebbles is $\approx2$~times higher than in M1 model. This is the consequence of the two pebble-depletion events in M1  model, that lead to 3 times lower pebble mass at 500\,kyr than the maximum value reached at 400\,kyr. The disc in M2 model did not experience such events by the end of the simulation, so its pebble mass is close to the maximum value. The difference in the pebble mass is directly translated to the difference in pebble ice masses. Overall, in terms of the ice chemical composition, these models are only slightly different, i.e., the mass of the initial cloud does not affect the main conclusions about the composition of ice on pebbles.

The pebble depletion events in the M1 model, which can be seen at about 400 and 460\,kyr in panels (b) and (f) of Figure~\ref{ris:2}, occur due to the episodic development of gravitational instability in the inner rings centred at 1-2 and 7-8~au. This is confirmed by the analysis of the Toomre Q-parameter, which drops below 1.5 during these events. These episodes of the ring instability are likely caused by mass influx from the colder disc regions beyond 10~au, where the gravitational instability persists throughout most of the considered disc evolution period. Normally, the rings represent local pressure maxima, towards which the grown dust drifts, separated by a pressure minimum in the gap between the two rings. When the gravitational instability develops in the rings, the gas density distribution is perturbed, and the gap between the two rings disappears.  As a result, pebbles that were localised in the outer ring begin to drift inward and cross the water snowline. Sublimation of water ice mantles lowers the fragmentation barrier of bare grains. Dust grains react to this new fragmentation barrier by decreasing in both the size and Stokes number below the values typical of pebbles. These grains are no longer able to regrow to pebbles because they are now in a warmer region and cannot sustain their ice mantles. A similar event occurs in the M2 model at later time (close to 500\,kyr).


\subsection{Two-dimensional distribution of pebbles}

\begin{figure*}
\center\includegraphics[width = 2\columnwidth]{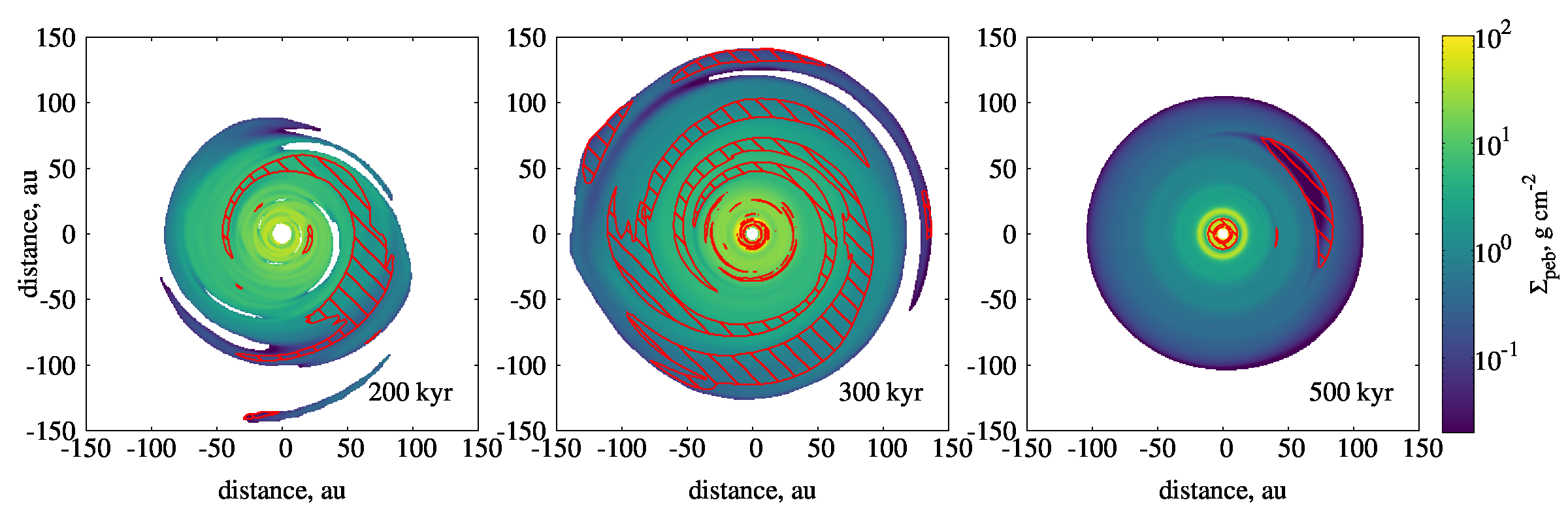}\\
\center\includegraphics[width = 2\columnwidth]{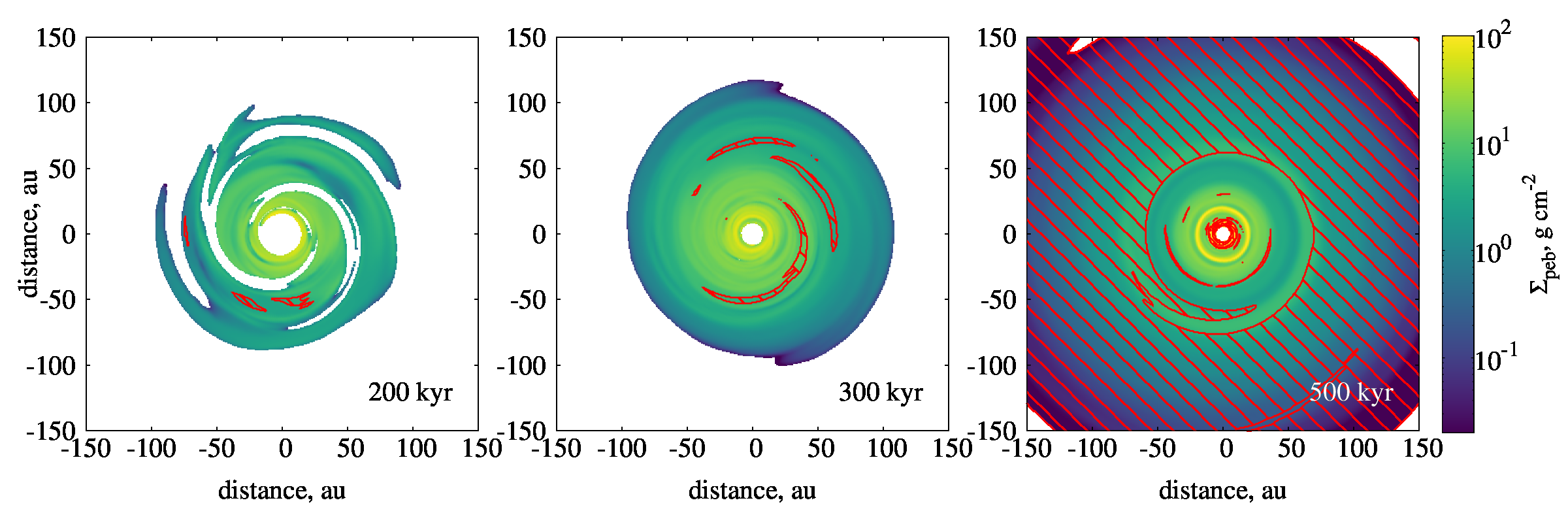}

\caption{Surface density of pebbles for M1 (top row) and M2 (bottom row) models at 200, 300, and 500\,kyr. The red contour shading show the result for pebble definition with $\mathrm{St}_{\rm 0}=0.05$.}
\label{ris:7}
\end{figure*}

Figure~\ref{ris:7} shows the surface density of pebbles in the protoplanetary disc at different times (200, 300, and 500\,kyr), which were calculated using the criteria from Section~\ref{sec:2B}. The upper panel shows the calculations for M1 model, and the lower panel is for M2 model. The surface density of pebbles decreases with time at the disc edge and increases closer to the central regions, for both cases. In the inner disc regions, at tens of au distances, $\Sigma_{\rm peb}$ is elevated due to dust growing and drifting to the star, and the formation of dust rings. At the early stages, the shapes of the pebble spatial distributions are asymmetric. The most noticeable difference between the models is that the area covered by pebbles in M2 model increases only by the 500\,kyr, while in M1 model this region reaches its maximum size earlier and decreases again by 500\,kyr. Total mass of pebbles in M1 model is also decreased by this time, which is related to the dynamics of the pebble-rich dust ring at $\approx$10\,au.

Figure~\ref{ris:7} also shows the region with pebbles if the lowest limit at Stokes number in the definition of pebbles is higher, $\mathrm{St}_{\rm 0}=0.05$. It can be seen that with this tighter constraint, the region of pebble existence is much smaller and has a fragmented shape. The two-dimensional maps for ices on pebbles are shown in Appendix~\ref{sec:apend}.

\subsection{Massive disc}
\label{sec:mass}

In order to check how total star plus disc mass affects the amount of pebbles formed in the disc, their composition and distribution, we performed calculations for a more massive model (M2 model in Table~\ref{tab:model}). Here we highlight the most important characteristics of pebbles in the massive model.

\begin{figure*} 
\center\includegraphics[width = 2\columnwidth]{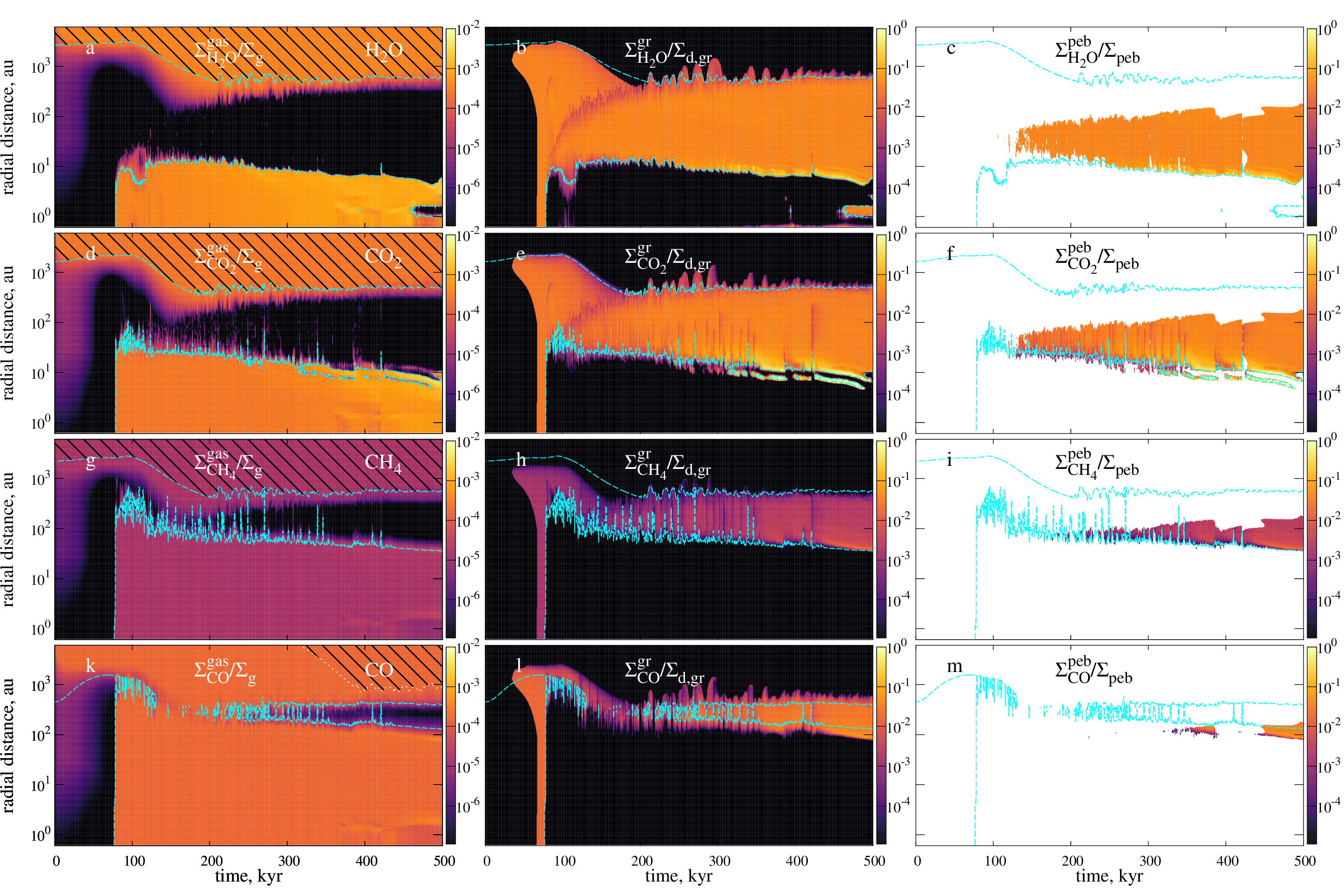} 
\caption{Evolution of the azimuthally averaged radial distributions of the four volatiles in the gas and in the ice on grown dust and on pebbles in M2 model. For each panel, the dashed cyan line shows the equilibrium position of the snowline.}
\label{ris:galkA_chem_1D}
\end{figure*}

Figure~\ref{ris:galkA_chem_1D} shows the evolution of the radial distributions of volatiles in M2 model. The dense dust ring at 1\,au in M2 model is formed later than in the reference model, it contains grown dust covered with water ice mantles, but the conditions for the existence of pebbles (Section~\ref{sec:2B}) are not fulfilled. In M2 model, ices of CO$_{2}$, CH$_{4}$, and CO appear on pebbles later than in the low-mass M1 model. The accumulation of volatiles in the ice phase in the vicinity of the snowlines is less pronounced, the annuli dominated by CH$_{4}$ and CO are absent, and the annuli dominated by CO$_{2}$ are weaker.
Otherwise, the composition of ices on pebbles in the models is similar, which is also shown in~Figure~\ref{ris:ices_total}.
In M2 model, H$_2$O and CO$_{2}$ are also the main volatiles on pebbles, so the conclusion of a reduced carbon abundance compared to ices on grown and small dust and volatiles in the gas remains valid.

The main difference between the two considered models is timing: the characteristic changes happen later in the more massive model. Unlike M1 model, where maximum pebble mass is reached at 400\,kyr, in M2 model, the pebble mass is growing until 500\,kyr. The events of sharp drops of pebble mass like those at 400 and 460\,kyr in M1 model do not occur in M2 model, until perhaps at $\approx 500$\,kyr, for which the consequences are unclear due to the limited time of calculation. Water ice in the ring at 1\,au appears 150\,kyr later. These differences in timing are caused by the differences in thermal structure. More massive disc is warmer, both because the central star providing the radiative heating is more massive and consequently hotter, and because the viscous heating is more efficient at higher gas surface densities. Thus the low enough temperatures in the inner disc are established later. Furthermore,  the snowlines of all species are further from the star, which is most notable for H$_2$O, as its snowline divides the regions with different $v_{\rm frag}$. The more massive M2 model reaches the thermal structure comparable to that of M1 model at later times.

Despite the fact that the total mass of gas and dust in M2 model is 1.5 times higher than in M1 model, the integrated pebble masses in the models are very similar. The shape and size of the main region where pebble exist in the two models are similar, although in the massive model the spiral structure in the shape of this region is more pronounced (see Figure~\ref{ris:7}). However, in M2 model the radial extent of this region gradually increases during almost the entire considered time interval, which can be traced from the H$_2$O ice distribution on pebbles (panel (c) in Figure~\ref{ris:galkA_chem_1D}). The inner edge of this region is a few au farther away from the star due to the farther position of the water snowline in this warmer disc. As the surface density of pebbles generally grows toward the centre, this lowers the total mass of pebbles in M2 model, making it close to that in M1 model.

\section{Discussion}
\label{sec:obs}

In the previous section, we analyse the amount of pebbles, their distribution and chemical composition of their ice mantles mostly focusing on a single protoplanetary disc model with certain parameters (M1 model). In this Section, we consider how particular model parameters can influence the properties of pebbles in the disc and our conclusions.

\subsection{Parameters of pebble definition}
\label{sec:criteria}

To distinguish pebbles from grown dust, we used the definition based on the work of \citet{Vorobyov2023A&A}, where the key parameters are the threshold values of $\mathrm{St}_{\rm 0}=0.01$ and $a_{\rm peb,0}=0.05$\,cm (see Section~\ref{sec:2B}). Here we also consider the following combinations of threshold parameters: $\mathrm{St}_{\rm 0}=0.01$ and $a_{\rm peb,0}=0.25$\,cm, $\mathrm{St}_{\rm 0}=0.05$ and $a_{\rm peb,0}=0.05$\,cm, and $\mathrm{St}_{\rm 0}=0.05$ and $a_{\rm peb,0}=0.25$\,cm. The chosen values are five times higher than those adopted in our reference model, so they correspond to larger and more dynamically decoupled pebbles and they imply a stricter restriction on the existence of pebbles.

\begin{figure}
\includegraphics[width = \columnwidth]{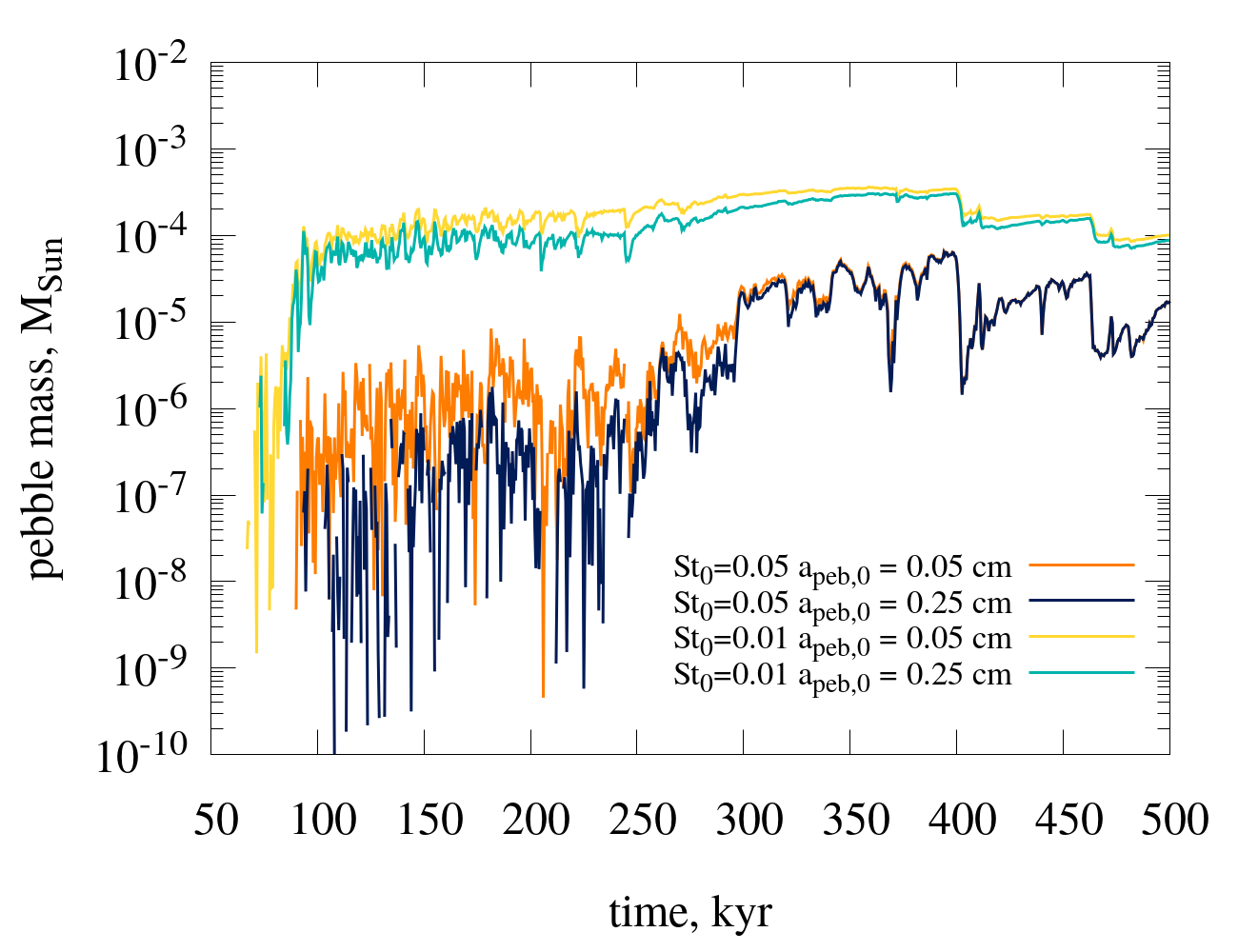}
\caption{Integrated mass of pebbles for different boundary values of the Stokes number $\mathrm{St}_{\rm 0}$ and the minimum possible pebble size $a_{\rm peb,0}$ in the definition of pebbles.} 
\label{fig:criteria_peb}
\end{figure}

The total mass of pebbles in the disc for the definition of pebbles with different parameters is shown in Figure~\ref{fig:criteria_peb}. If we compare the cases of different $a_{\rm peb,0}$ at fixed $\mathrm{St}_{\rm 0}$, we see that changing $a_{\rm peb,0}$ leads to an order of unity difference in the total pebble mass. At $\mathrm{St}_{\rm 0}=0.05$, changing $a_{\rm peb,0}$ has almost no effect on the pebble mass after 300\,kyr, i.e., all pebbles are $>0.25$\,cm.

If we change the critical Stokes number, the pebble mass in the disc drops significantly. The main effect of applying more strict definition of pebbles is the reduced areas where pebbles exist and lower surface density of pebbles. The former has already been shown in the two-dimensional distributions in Figure~\ref{ris:7}, where the shaded areas ($\mathrm{St}_{\rm 0}=0.05$) are much smaller compared to the coloured ones ($\mathrm{St}_{\rm 0}=0.01$). The value of $\Sigma_{\rm peb}$ in these areas is also generally lower, because $a_{\rm peb,min}$ depends on $\mathrm{St}_{\rm 0}$ (see Eqs.~\eqref{eq:apebmin} and~\eqref{eq:sigmapeb}). With $\mathrm{St}_{\rm 0}=0.05$, the total mass of pebbles does not exceed $10^{-5}$\,$M_{\odot}$ until almost 300\,kyr. The maximum value of $6.4 \times10^{-5}$\,$M_{\odot}$ is $\approx21$\,$M_{\oplus}$, which is a factor of 5 smaller than with the baseline definition. Thus, most pebbles in the model have Stokes number in the range $0.01<\mathrm{St}<0.05$. Stokes parameter is the most important boundary condition in the definition of pebbles.

As can be seen from panel (h) of Figure~\ref{ris:2}, grains with $\mathrm{St}>0.1$ are practically absent in the protoplanetary disc. Many models of planet formation through pebble accretion consider high values of $\mathrm{St}$, such as 0.1 or 1 \citep[e.g.,][]{Lambrechts2012}. We show that grains with such $\mathrm{St}$ values are hardly available in young discs with dust growth self-consistently formed through the collapse of a prestellar core.


\subsection{Chemical composition of the ice}
\label{sec:ices_initial}

The present work is focused on the composition of the ice mantles on pebbles in the context of global dust growth and migration in a protoplanetary discs. We do not aim to accurately predict the composition of ices, and rather describe it in the first-order approximation. There are many astrochemical models of protoplanetary discs that employ comprehensive chemical reaction networks, such as ALCHEMIC \citep{2010A&A...522A..42S}, ANDES~\citep{2013ApJ...766....8A}, NAUTILUS \citep{2015ApJS..217...20W} or ProDiMo \citep{2017A&A...607A..41K}. They include gas-phase and surface reactions and various sources of ionisation. However, these models do not describe global gas and dust dynamics, which is the focus of FEOSAD code. There are many other chemical models of protoplanetary discs that include different other physical processes and approximations, see the review by \citet{2023ARA&A..61..287O} and references therein. Some hydrodynamic simulations of protoplanetary discs include complicated chemical kinetics modelling \citep{2017MNRAS.472..189I,2019ApJ...874...90W}. However, they only allow to follow disc evolution for relatively short timescales ($\sim 10^3$~yr). In FEOSAD, we describe global evolution of a dusty disc, beginning from the core collapse, and therefore need to use a simplified chemical model allowing for fast analytical solution.

The initial abundances of the volatiles should influence the final composition of ice on pebbles. Apart from the four most dominant volatiles, other species, including complex organics \citep{2012A&A...541L..12B,2022arXiv220613270C}, are also present in the composition of protostellar cores and are not considered in the present paper. The abundances of these species are much lower than water, carbon dioxide, or carbon monoxide, but together they can contribute significantly to the carbon and oxygen balance of the volatiles. 

The adopted ice fractions are based on the observed ice composition in protostellar cores \citep{Oberg2011a}. The mass fraction of the ices relative to the total gas mass was calculated under the assumption that the water abundance relative to the gas number density is $5\times10^{-5}$, which is based on numerical models of \citet{2004A&A...426..925P, 2004ASPC..309..547B}. However, under these assumptions, the total mass of ice in the model appears to be more than an order of magnitude lower than the mass of the refractory component ($8.5\%$). Data from the Solar System studies and the observations of the ISM suggest that dust grains in protoplanetary discs should be covered by ice mantles with masses comparable to the refractory dust grain cores, or even exceeding them by up to $2-4$ times~\citep{2014prpl.conf..363P}. In line with this, for example, modelling the formation of planetesimals on the water snowline, \citet{2017A&A...608A..92D} assumed an initial water content of $50\%$ of the dust mass, implying equal masses of water mantles and refractory cores. Using this higher abundance of water could affect the ratio between the volatile and refractory components of dust grains in our modelling results.

The choice of the low initial mass fraction of ice is also justified by the model limitations. We neglect the contribution of ice mantles to the mass and size of dust grains and assume that mantles do not affect their dynamics. Since the model implies zero-order desorption, the local amount of a volatile species in the ice phase also does not affect the desorption rate, but the adsorption (freeze-out) rate is proportional to the current amount of the gas-phase volatile species. Thus, an increase in the abundance of volatiles by several times should lead to a shift of the snowlines closer to the star, i.e., to the expansion of the ice phase dominated region.

As the first approximation, we can claim that the resulting surface densities of ices in the disc, including ices on pebbles, are proportional to the initial mass fractions of the corresponding volatiles. However, one should remember that the assumption of massless ice mantles will no longer hold with the increased initial ice masses. With the  mass fractions of volatiles adopted in the presented modelling, masses of ice mantles exceed masses of refractory cores in some parts of the disc. With higher initial ice fractions, this assumption might be violated in the whole disc. The modelling of adsorption and desorption of mantles for dust grains mainly consisting of volatiles is a problem on its own relevant for the conditions of protoplanetary discs and it merits a focused consideration.

During protoplanetary disc lifetimes, the chemical composition of ice and gas will also inevitably change due to chemical reactions in the gas phase and on the surface of the dust grains~\citep{2013ChRv..113.9016H}. These processes include surface chemical reactions \citep{1992ApJS...82..167H,2008A&A...491..239G}, photo-reactions in ice mantles under UV irradiation \citep{2009A&A...504..891O} and sublimation by cosmic rays \citep{2015ApJ...805...59I}.
This would lead to the formation of more complex species with different desorption energies, which could transfer to the other phase and hence participate in the motion of the other dynamical component (grown dust or gas/small dust). In the presented model we consider only the most important chemical processes: adsorption and desorption, but the role of other chemical reactions may also be decisive. Modelling dust drift in a viscous disc in combination with a large chemical reaction network, \citet{2019MNRAS.487.3998B} demonstrated that the direct transport of chemical species with dust grains dominates over chemical reactions, determining the distribution of volatiles, if $\alpha$ is high enough ($>10^{-3}$). In our model with variable $\alpha$-parameter, this condition is fulfilled outside the dead zone ($\approx10-20$\,au), however, in the inner regions the role of chemistry may be more important and should be investigated in separate studies.

The desorption energies we adopted from \citet{Molyarova2021ApJ} can also be model parameters. In particular, they depend on the presence of other ices within the mantle \citep{2017SSRv..212....1C}. Changes in the desorption energies can lead to shifts in the snowlines. For example, for CO, a shift of the snowline by $\approx 10$\,au can lead to this species completely absent on pebbles and prevent the formation of a region where CO dominates the ice composition (see Section~\ref{sec:solid_composition} and Figure~\ref{ris:3}). Another important parameter is the value of pre-exponential factor. As shown by \citet{2022ESC.....6..597M}, the widely used approach by \citet{1993MNRAS.263..589H}, which we also use, underestimates pre-exponential factor. It produces values that are $3-5$ orders of magnitude lower compared to both experimental data and the transition state theory approach, that give similar values. Using those newer values would allow to more precisely determine the positions of the snowlines in protoplanetary discs.

To test the effect of the chosen thermal desorption parameters on our results, we calculated the equilibrium positions of the snowlines in M1~model using the recommended values of desorption energy and pre-exponential factor from Table~5 of~\citet{2022ESC.....6..597M}. For H$_2$O, the positions of the snowlines changed only slightly. The temperature gradient is high around the water snowline, so even though the freeze-out temperature is significantly different, the position is almost the same. For CH$_4$, the difference in the parameters is relatively small, and the snowline is shifted inwards by $\sim$10\,au. The snowline of CO$_2$ is shifted outwards by $3-5$\,au. Its new position is beyond the dust ring at $\sim10$\,au, and the region with CO$_2$ ice inside 10\,au disappears. For CO, the snowline is shifted significantly mostly due to the change in binding energy, which is a factor of~$1.5$ higher in~\citet{2022ESC.....6..597M}. However, the CO desorption energies previously discussed in the literature~\citep[e.g.][]{2017SSRv..212....1C} are not as high. At the same time, if only the pre-exponential factor is to be changed, then no CO ice is present in the disc at all. This is an ambiguous result, which hampers the determination of the CO snowline position, as it is unclear what desorption parameters are better to use.

\subsection{Fragmentation velocity}
\label{sec:vfrag_discussion}

The surface density of ice on grown dust in the model has a direct effect on the fragmentation velocity. However, the threshold value of the amount of ice required to cover a dust grain with one monolayer and thus to affect the value of $v_{\rm frag}$ is very low, on the order of $10^{-5}\times \Sigma_{\rm d, gr}$ for the millimetre-sized dust \citep[see Equation (40) in Section 2.6 of][]{Molyarova2021ApJ}. A proportional increase in the amount of ice means that this condition will be satisfied in a slightly larger part of the disc. Given that in the vicinity of the snowline the surface density of volatiles changes exponentially in the radial direction, the region where dust grains are covered by ice will not change significantly.

More significant changes should be expected when considering other values of $v_{\rm frag}$. The values of $v_{\rm frag}=0.5$ and 5\,m~s$^{-1}$ adopted in this paper are relatively low. For example, in \citet{Molyarova2021ApJ}, three times higher fragmentation velocities were considered. Numerical modelling shows that ice-covered dust aggregates can avoid fragmentation in collisions at velocities up to 50\,m~s$^{-1}$~\citep{2009ApJ...702.1490W}. Meanwhile, in laboratory experiments, dust grains with ice mantles fragment at velocities above $10-15$\,m~s$^{-1}$~\citep{2015ApJ...798...34G}. However, $v_{\rm frag}$ also depends on the monomer size, which in our model is $a_{*}=1$\,$\mu$m, so a lower value of fragmentation velocity seems more appropriate~\citep[see Eqs.~(3) and~(4) in][and related references]{2019ApJ...878..132O}. The choice of higher values of $v_{\rm frag}$ weakens the role of fragmentation and allows dust grains to coagulate over a wider range of conditions. However, due to dust radial drift, this does not guarantee the survival of these grown dust grains in the disc~\citep{1977MNRAS.180...57W}.

At higher $v_{\rm frag}$, we can expect an increase in the total amount of pebbles in the disc, but their accretion rate onto the star should also increase: in this situation, the relative fraction of solids of the disc can become lower. In \citet{2020A&A...637A...5E}, a higher fragmentation threshold of $30$\,m~s$^{-1}$ is adopted, and dust grains reach a size of a few centimetres. Their comparison of models with fixed $\alpha=0.01$ and different $v_{\rm frag}=1$ and 30\,m~s$^{-1}$ shows that lower fragmentation velocity leads to a smaller dust size (of the order of millimetres). However, the global disc evolution is weakly affected because at $\alpha=0.01$ the drift is determined by advection and dust is drifting with gas regardless of its size. For lower $\alpha$ values, the situation must be different when such a broad span in $v_{\rm frag}$ is considered. The fact that the amount of pebbles in the disc is very sensitive to $v_{\rm frag}$ for $\alpha\leq10^{-3}$ was thoroughly demonstrated in \citet{Vorobyov2023A&A}.

If we discard the considered effect of ice mantles and assume a relatively high value of $v_{\rm frag}$ for regions with ``bare'' silicate dust grains, then pebbles without ice mantles may also exist in the disc. Another caveat of the model is that  the chemical composition of the mantle does not affect the value of fragmentation velocity. In particular, we do not consider the effect of CO$_2$, which in laboratory experiments was shown to reduce $v_{\rm frag}$ to values typical of bare silicate dust \citep{2016ApJ...818...16M, 2016ApJ...827...63M}.

\section{Conclusions}
\label{sec:vivod}

In this work, we model the formation and global evolution of a self-gravitating viscous protoplanetary disc using the two-dimensional hydrodynamic code FEOSAD. The simulations include gas dynamics, dust dynamics and growth, thermal balance, and volatile dynamics starting from cloud collapse to 500\,kyr. We analyse the distribution of pebbles and the distribution of the main volatile species (H$_2$O, CO$_{2}$, CH$_{4}$ and CO) in the gas, on small dust, on grown dust and on pebbles. The main conclusions of the work are listed below. 

\begin{itemize}
    \item We show that pebbles are present in a protoplanetary disc already $\approx50$\,kyr after the disc formation until the end of modelling (0.5\,Myr). The integral mass of pebbles in the disc reaches a value of $10^{-4} M_{\odot}$ at $\approx50$\,kyr, subsequently increasing by a factor of three and falling back to $10^{-4} M_{\odot}$ by the end of the simulation, which is from a few percent to a quarter of the mass of all grown dust. The maximum value of the total pebble mass (excluding ice mantles) is about $3.5\times 10^{-4} M_{\odot}$ or $\approx115$\,$M_{\oplus}$, which is comparable to the mass of solid matter in the Minimum Mass Solar Nebula. This suggests that there are prerequisites for planet formation at the early stages of disc evolution.
    \item All pebbles present in the disc are covered by ice mantles, i.e., there are no ``bare'' pebbles consisting exclusively of the refractory core. Ice mantles on pebbles  consist mainly of H$_{2}$O and CO$_{2}$, and are carbon-depleted compared to the mantles on small dust, grown dust, and to the gas, which contain more CO$_{2}$ and CH$_{4}$. This suggests a possible dominance of oxygen in the composition of planets formed from pebbles under these conditions.

    \item Most of the pebbles in the model have Stokes number between 0.01 and 0.05, and the grains with $\mathrm{St}>0.1$ are practically absent in the disc, which should be accounted for in the planet formation models.
\end{itemize}

We consider different parameters in the definition of pebbles, as well as two models with different initial masses of the collapsing cloud. Comparison of the results showed that the above conclusions stay valid for the considered model parameters.

\section*{Acknowledgements}

We are grateful to Vardan Elbakyan for fruitful discussions regarding the definition of pebbles. The study was funded by a grant from the Russian Science Foundation 22-72-10029, https://rscf.ru/project/22-72-10029/

\section*{Data availability}

The data underlying this article will be shared on reasonable request to the corresponding author.

\bibliographystyle{mnras}
\bibliography{mnras_template}

\appendix

\section{Different criteria for pebble definition}
\label{sec:pebble_appendix}

In this paper we use a modified version of the definition of pebbles from \citet{Vorobyov2023A&A}. The main difference is in the definition of $a_{\rm peb,min}$, which is defined in Eq.~(27) in \citet{Vorobyov2023A&A} and in Eq.~\eqref{eq:apebmin} in the present study. If in some disc point, $a_{\rm St_0}$ is below $a_{\rm peb,0}$~\citep[0.01 and 0.05\,cm, in][respectively]{Vorobyov2023A&A}, then in the previous definition $a_{\rm peb,min}$ simply goes to zero. However, there still exist grains with higher $\mathrm{St}$, which are larger that $a_{\rm peb,0}$ in this point, which are excluded from consideration by this definition. Instead of neglecting all dust in this point, we set higher value of $a_{\rm peb,min}$, which allows us to account for more grains large enough to be called pebbles.

Another advantage of the modified definition of pebbles is that it allows to more consistently compare the different versions of the boundary value $\mathrm{St_0}$. In the previous definition, the regions discarded at $\mathrm{St_0}=0.01$ could be included at $\mathrm{St_0}=0.05$, if the corresponding grain sizes $a_{\rm St=0.01}$ and $a_{\rm St=0.05}$ were on the opposite sides of $a_{\rm peb,0}=0.05$\,cm, i.e., $a_{\rm St=0.01}<a_{\rm peb,0}<a_{\rm  St=0.05}$. So at different $\mathrm{St_0}$ the regions where pebble exist would be independent. In the modified definition, the regions with higher $\mathrm{St_0}$ or $a_{\rm peb,min}$ are always the subset of the regions with lower boundary values.

\begin{figure} 
\includegraphics[width = \columnwidth]{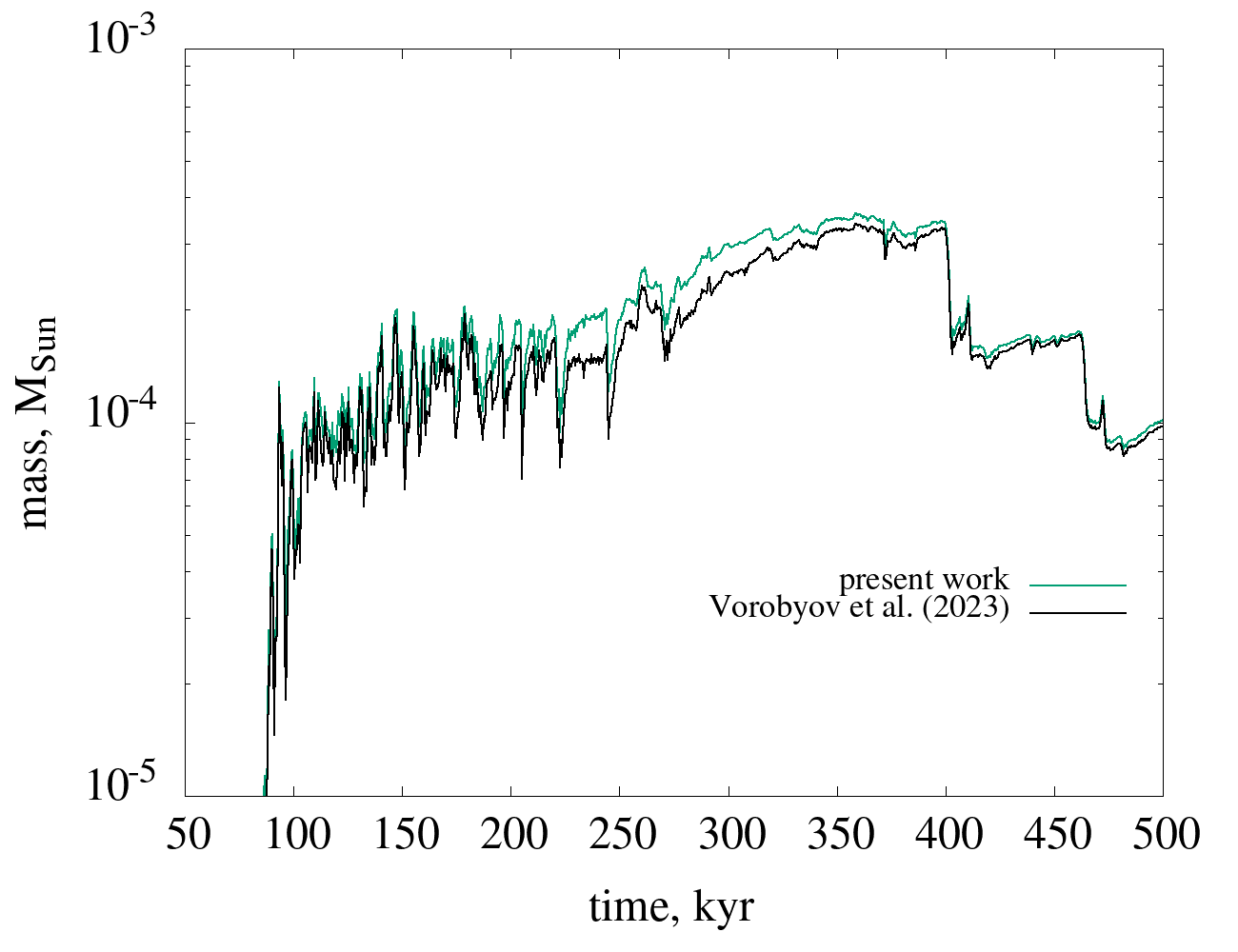}
\caption{Integrated mass of pebbles for different definitions of pebbles. The adopted definition is shown in green, the definition of~\citet{Vorobyov2023A&A} is in black.}
\label{ris:oldnew}
\end{figure}

Figure~\ref{ris:oldnew} shows integrated mass of pebbles for these two different definitions of pebbles with the same boundary values $\mathrm{St_0}=0.01$ and $a_{\rm peb,0}=0.05$\,cm. Between 200 and 400\,kyr the values of pebble masses are different by a factor of unity, but earlier and later the difference is very small. This means that the contribution of the regions discarded by the definition of~\citet{Vorobyov2023A&A} is not large in terms of total pebble amount. However, we believe that it is important to use the new definition that includes pebbles from more disc regions. The formation of planets may already begin at 200\,kyr, and the presence of pebbles in wider range of conditions may affect the rate of formation of planets. Moreover, the composition of ices in the discarded regions can also be different and relevant for the composition of planets.

\section{Two-dimensional distribution of ices on pebbles}
\label{sec:apend}

\begin{figure*} 
\includegraphics[width = 7cm]{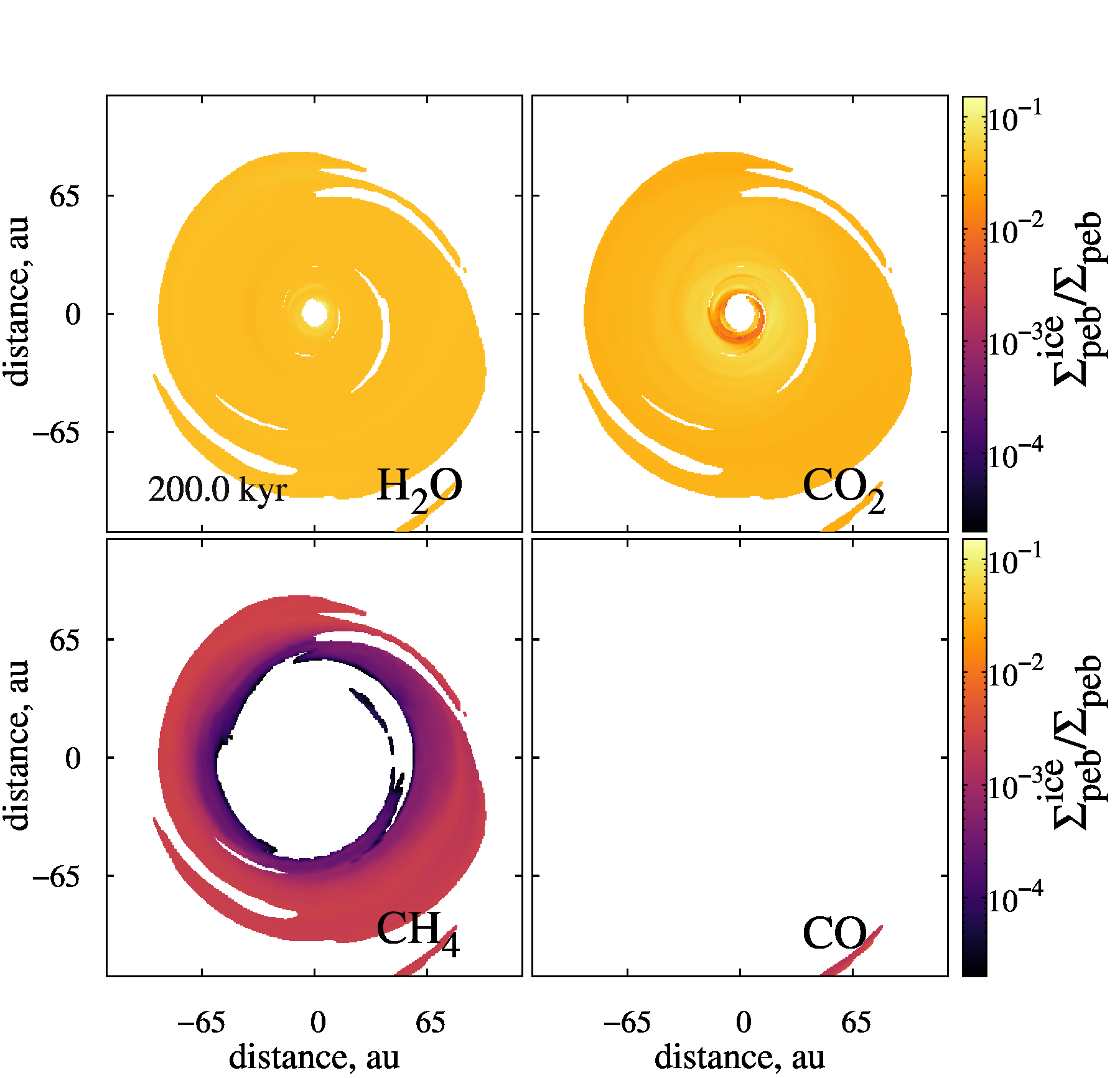}
\includegraphics[width = 7cm]{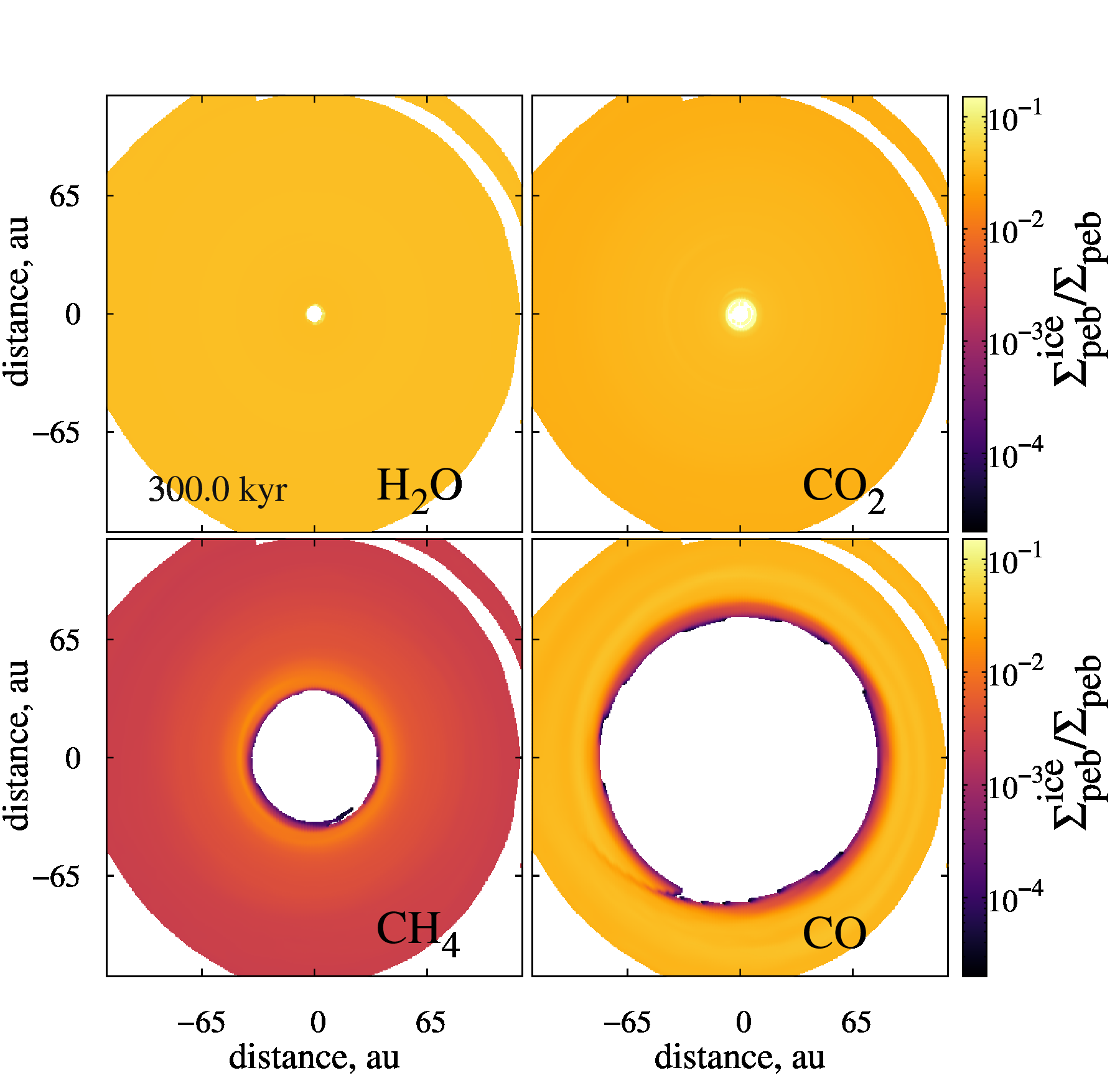}
\includegraphics[width = 7cm]{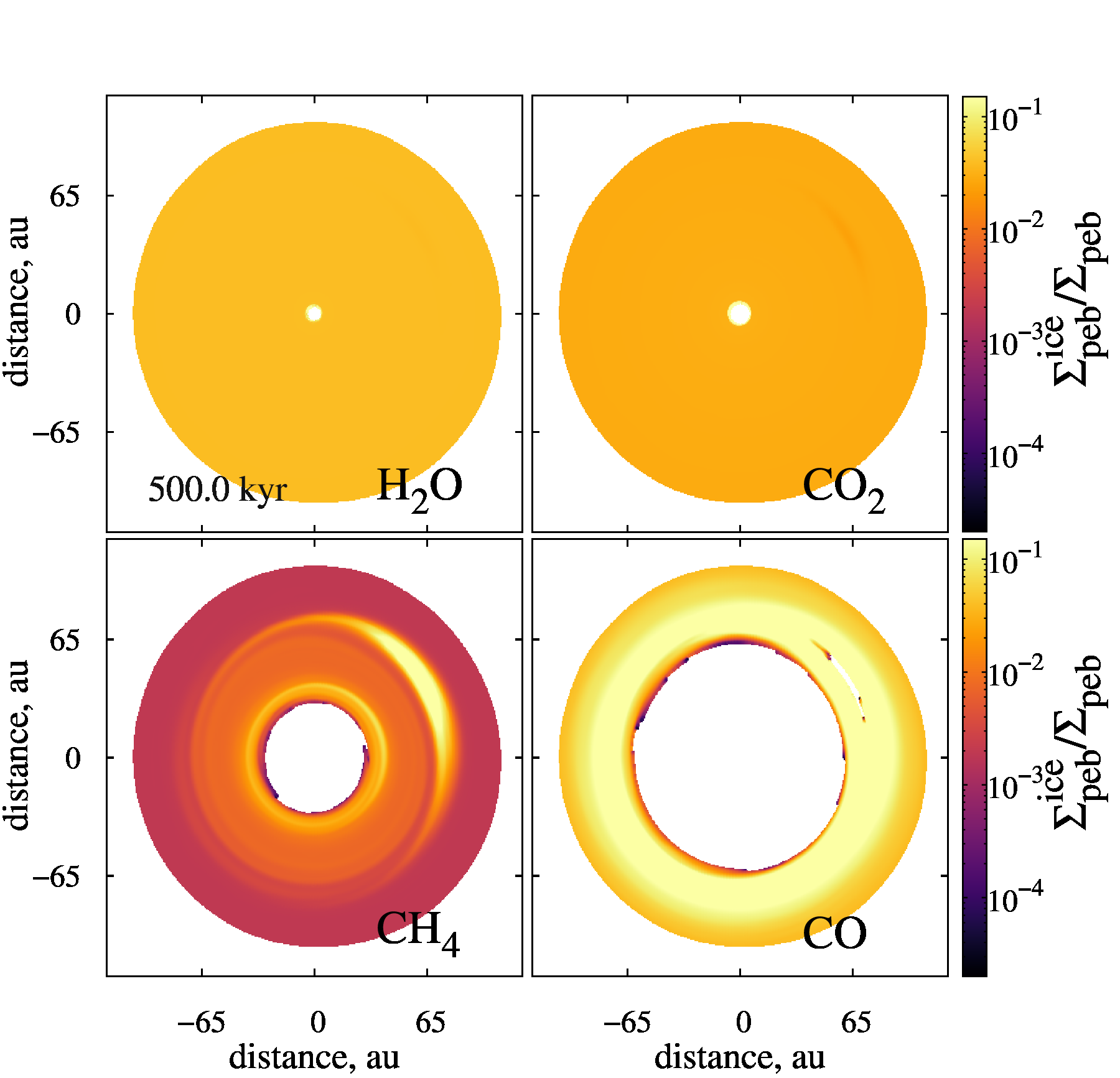}
\caption{Mass fraction of H$_{2}$O, CO$_{2}$, CH$_{4}$ and CO ice on pebbles in M1 model at 200, 300, and 500\,kyr.}
\label{ris:44}
\end{figure*}

Figure~\ref{ris:44} shows the surface densities of H$_{2}$O, CO$_{2}$, CH$_{4}$, and CO ice on pebbles relative to the surface density of pebbles in M1 model at different times (200, 300, and 500\,kyr). The distribution of H$_{2}$O repeats the distribution of pebbles in Figure~\ref{ris:7}. It can also be seen that H$_{2}$O and CO$_{2}$ ices are the most abundant on pebbles, and they are evenly distributed over the disc.

\bsp	
\label{lastpage}
\end{document}